\documentstyle[preprint,aps,prd]{revtex}
\tightenlines
\begin{document}
\newcommand{\eq}{\begin{eqnarray}}
\newcommand{\en}{\end{eqnarray}}
\newcommand{\Ms}{M^\star}
\newcommand{\Ps}{P^\star}
\newcommand{\Es}{E^\star}
\newcommand{\mc}{m_\pi^2}
\newcommand{\mo}{m_{\pi^0}^2}
\newcommand{\dpi}{\Delta_\pi}
\draft
\title{$\pi^+\pi^-$ Atom in Chiral Perturbation Theory}

\author{M.\ A.\ Ivanov}
\address{Bogoliubov Laboratory of Theoretical Physics, Joint Institute
for Nuclear Research, 141980 Dubna, Russia}

\author{V.\ E.\ Lyubovitskij}
\address{Bogoliubov Laboratory of Theoretical Physics, Joint Institute
for Nuclear Research, 141980 Dubna, Russia
and Department of Physics, Tomsk State University,
634050 Tomsk, Russia}

\author{E.\ Z.\ Lipartia}
\address{Laboratory for Computational Technique and Automation,
Joint Institute for Nuclear Research, 141980 Dubna, Russia
and IHEP, Tbilisi State University, 380086 Tbilisi, Georgia}

\author{A.\ G.\ Rusetsky}
\address{Bogoliubov Laboratory of Theoretical Physics, Joint Institute
for Nuclear Research, 141980 Dubna, Russia
and IHEP, Tbilisi State University, 380086 Tbilisi, Georgia}

\date{\today}

\maketitle

\begin{abstract}\widetext
Hadronic $\pi^+\pi^-$ atom is studied in the relativistic perturbative
approach based on the Bethe-Salpeter equation. The general expression for
the atom lifetime is derived. Lowest-order corrections to the relativistic
Deser-type formula for the atom lifetime are evaluated within the
Chiral Perturbation Theory.
\end{abstract}


\section{Introduction}

The pion-pion scattering amplitude at low energies forms one of the basic
building blocks in the hierarchy of strong interaction processes.
It serves as a useful probe for the investigation of the effect of
Chiral Symmetry breaking since in the chiral limit the pion interactions
vanish at the threshold. According to common belief, low-energy interactions
of pions are described within the Chiral Perturbation Theory (ChPT)
\cite{ChPT,Stern} which exploits the full content of global
QCD symmetries. The $\pi\pi\to\pi\pi$ amplitude in the ChPT is obtained
as an expansion in quark masses and external pion momenta.
The predictions of ChPT are sensitive to the magnitude of the quark condensate
$<0|\bar q q|0>$. In the standard scheme \cite{ChPT}
with a "large" condensate, the $S$-wave $\pi\pi$ scattering
lengths are predicted
to be $a_0^0=0.217$ and $a_0^0-a_0^2=0.258$ in units of the inverse
charged pion mass \cite{Bijnens}. The calculations within the Generalized
ChPT with a small quark condensate yield $a_0^0=0.27$ \cite{Stern}.
Despite a significant difference between these predictions, both
results for the scattering length $a_0^0$ are compatible with the
experimental value $a_0^0=0.26\pm 0.5$ \cite{Rosselet}. Consequently, a
precise measurement of $\pi\pi$ scattering lengths will be an excellent
test of the ChPT. In particular, an experiment of that sort would provide an
important information about the behavior of the quark condensate in the
chiral limit which in its turn is related to the properties of the QCD gluon
vacuum.

The experimental study of the $\pi\pi$ scattering process is a very
difficult task mainly due to the absence of the pion target.
Indirect information extracted from the available data for, e.g.,
the process $\pi N\rightarrow\pi\pi N$ \cite{pNppN}, produces large error
bars when extrapolated to the two-pion threshold. The study of $K_{e4}$
decay which is preferable for determining of the parameters of
$\pi\pi$ interaction near threshold, is complicated because of a very small
branching ratio of this process \cite{Rosselet} (for the review of the
recent status of $\pi\pi$ experiments see, e.g. Refs. \cite{Pocanic}).
In view of the above-mentioned experimental uncertainty in the determination
of the $\pi\pi$ data near threshold, the forthcoming high-precision
measurement of the $\pi^+\pi^-$ atom lifetime by the DIRAC collaboration
at CERN (project \# PS212) attracts much attention since it will allow
the direct determination of the difference $a_0^0-a_0^2$ and thus will
provide an excellent probe for the predictions of ChPT. The possibility
of observation of such atoms was argued in Ref. \cite{Nemenov}.
The first estimation of the lifetime of an atom formed by $\pi^+$ and
$\pi^-$ in the ground $1S$ state
$\tau_1=2.9^{+\infty}_{-2.1}\times 10^{-15}~s$ was given
in Refs. \cite{Afanasyev}. The expected high-precision experimental data
from DIRAC experiment call for a refined theoretical
treatment of this sort of bound systems.

Since the characteristic average momentum in hadronic atoms are
of an order of a few MeV, these systems are highly nonrelativistic.
With the use of this fact the nonrelativistic Deser formula
was derived in Refs. \cite{Deser,Uretsky}. For the particular case
of the $\pi^+\pi^-$ atom the formula reads
\eq\label{DESER}
\tau_n^{-1}=\frac{16\pi}{9}\,
\biggl(\frac{2\,\Delta m_\pi}{m_\pi}\biggr)^{{1}/{2}}
(a_0^0-a_0^2)^2\,\,|\Psi_n(0)|^2
\en
\noindent
which relates the lifetime $\tau_n$ of the atom in the $n$-excited state
to the value of the Coulomb wave function (w.f.) of the atom at the
origin $\Psi_n(0)$ and the difference of the $S$-wave $\pi\pi$ scattering
lengths with the total isospin $I=0$ and $I=2$. Note that in (\ref{DESER})
one assumes the isospin symmetry when expressing the scattering length
for the reaction $\pi^+\pi^-\rightarrow\pi^0\pi^0$ in terms of scattering
lengths with a definite total isospin, though the factor $\Delta m_\pi$
in the r.h.s. of Eq. (\ref{DESER}) indicates the
necessity of taking into account the isospin-breaking effects in the
theoretical description of the decay process. Note also that the Deser-type
formulae for the energy level displacement and lifetime are now widely
used for the theoretical analysis of the experimental data for other
hydrogenlike bound systems, such as pionic hydrogen \cite{Sigg},
pionic deuterium \cite{Chattelard}, etc.

It is well known that the Deser-type formulae for hadronic atoms
in the nonrelativistic scattering theory are valid up to the electromagnetic
corrections to hadron scattering processes, if the mass difference
between the charged and neutral components of the isotopic multiplet
is assumed to be of a purely electromagnetic origin. Put differently,
the clear-cut factorization of strong and electromagnetic interactions
in the hadronic atom observables, which is explicit in Eq. (\ref{DESER}),
is valid up to (small) electromagnetic effects. In Ref. \cite{Trueman}
the regular approach was constructed for the evaluation of these corrections.
The ideas of this approach have been successfully applied to the study of
the properties of a $\pi^+\pi^-$ atom in Refs. \cite{Rasche1,Rasche2} where the
coupled-channel scattering problem with ($\pi^+\pi^-$) and ($\pi^0\pi^0$)
free pairs in the asymptotic states has been considered.
The nonrelativistic scattering theory has been used for the
investigation of hadronic atoms also in Refs. \cite{Gibbs,Geramb}.
Note, however, that very "narrow" and "deep" phenomenological $\pi\pi$
potentials used in Ref. \cite{Geramb} lead to the instability in the
calculated observables of the bound state with respect to a small variation
of input parameters. A strong enhancement of the potential in the vicinity
of the origin within the inverse scattering theory approach, used by the
authors of Ref. \cite{Geramb}, stems mainly from the particular
parameterization of the $\pi\pi$ scattering phase shift in the high energy
domain, where theoretical calculations of this quantity cannot be
performed.

To summarize, the lowest-order Deser formula (\ref{DESER}) and its
counterpart for the energy-level displacement of an atom due to strong
interactions are valid irrespective of the concrete choice of the
strong interaction potential between hadrons. The magnitude, and even
the sign of corrections to it, however, strongly depend on this choice.
For the theoretical analysis of the high-precision experimental data expected
from DIRAC experiment, the model-independent evaluation of these
corrections is needed, based on the underlying (chiral) Lagrangian
dynamics of hadrons, rather than the nonrelativistic potential picture
of strong interactions.

The problem of relativistic description of hadronic atoms is much
richer in content than the same problem in the nonrelativistic scattering
theory formulation. Many new effects which were absent, or were mimicked
in the potential treatment, now arise naturally from the beginning (e.g.,
the "vacuum polarization" and "finite size" corrections which are borrowed
in the potential picture from field theory). Namely, the problem of
evaluation of the atom lifetime on the basis of the underlying
strong interaction dynamics was addressed to in Refs. \cite{Efimov,Pervushin}.
Ref. \cite{Silagadze} deals with the so-called retardation correction
in the pionium lifetime. In Ref. \cite{Kuraev} the radiative corrections
to $\pi\pi$ scattering lengths have been evaluated, which induce
the corresponding $O(\alpha)$ correction in the pionium lifetime.
Recently, the paper by H.~Jallouli and H. Sazdjian appeared \cite{Sazdjian}
which is aimed at the consistent description of the properties
of $\pi^+\pi^-$ atom on the basis of the 3D bound-state
equation obtained in the framework of constraint theory, with the underlying
strong $\pi\pi$ interactions described by ChPT. The authors have calculated
the corrections to the pionium lifetime coming from the mass difference
$m_{\pi^\pm}-m_{\pi^0}$, as well as the corrections from second-order
perturbation theory and electromagnetic radiative corrections.
In paper \cite{Labelle} the correction due to vacuum polarization to
the pionium lifetime was calculated. Below we
shall present the detailed comparison of our results with those given
in Refs. \cite{Sazdjian,Labelle}.

Our previous papers \cite{PLB,JETP} were aimed at the consistent
field-theoretical treatment of $\pi^+\pi^-$ atom observables on the basis
of the Bethe-Salpeter (BS) equation. Namely, in Ref. \cite{PLB}
we derived the relativistic analogue of the Deser formula (\ref{DESER})
for the pionium lifetime and evaluated the correction to it, coming
from the displacement of the bound-state pole by strong interactions,
referred hereafter to as "strong correction".
In Ref. \cite{JETP} we presented a systematic perturbative approach
based on the BS equation for the calculation of hadronic atom observables.
In this paper we give a closed expression, containing all first-order
corrections to the pionium lifetime, and evaluate part of them. Namely, apart
from the "strong correction", which is reproduced here, we calculate the
correction due to the exchange of Coulomb photon ladders, corresponding
to the "second-order perturbation theory" correction from Ref. \cite{Sazdjian},
and the relativistic correction to the bound-state w.f.

The purpose of the present paper is to give a detailed description
of the systematic perturbative approach to the hadronic atom characteristics,
based on the BS bound-state equation. This approach is by construction free of
any double-counting problems.
Within this approach we collect together and
calculate or give an estimate of all lowest-order corrections
to the Deser-type relativistic formula for the pionium lifetime.
The underlying strong $\pi\pi$ interactions within our approach are
described in ChPT. Consequently, the results of present calculations
of the lifetime of a $\pi^+\pi^-$ atom form a self-consistent
basis for the verification of the predictions of ChPT in
the DIRAC experiment. Further, in the present paper we discuss in detail
the links to other approaches used for the description of a $\pi^+\pi^-$
atom and, in particular, that from Refs. \cite{Sazdjian,Labelle}.

The layout of the present paper is follows: In Sect. II, we present
a detailed description of the perturbative approach to the $\pi^+\pi^-$
atom characteristics. In this section, we give a closed expression
for the first-order correction to the pionium lifetime. In Sect. III,
we give an evaluation, term by term, of various first-order corrections
to the Deser formula (\ref{DESER}). Sect. IV contains our conclusions.

\section{Perturbative Bethe-Salpeter approach to $\pi^+\pi^-$ atom}

The evaluation of corrections to the Deser-type formulae for
hadronic atom observables cannot be confined solely to the evaluation
of corrections to the pionium w.f. or to scattering lengths in
Eq. (\ref{DESER}). One has to develop a consistent perturbative scheme
for the calculation of atom observables, which in the lowest-order
approximation should yield the Deser-type relations for these quantities.
Our approach is based on the field-theoretical BS
equation with the kernel constructed from the underlying Lagrangian
of ChPT.

Below we shall briefly discuss the basic ideas and assumptions
of the approach which employs the following physical picture.
The formation of a $\pi^+\pi^-$ atom proceeds mainly due to the static
Coulomb potential whereas strong interactions are responsible for its decay.
The atom is described by the exact w.f. which obeys the field
theoretical BS equation. For our purposes we split this
kernel into the Coulomb piece and the remainder, the latter considered in our
scheme as a perturbation. Then, the exact BS w.f. is related to the
relativistic Coulomb w.f. in the perturbation theory. The crucial point
of our approach is that with the use of the above relation the observables
of an atom (lifetime and binding energy) in every perturbative order
are expressed in terms of the Coulomb w.f. In the leading order
of a perturbative expansion we reproduce the Deser formulae for atom
observables \cite{Deser}. The next-to-leading term in the perturbative
expansion produces all lowest-order corrections to the Deser formulae.

Let us now pass to the description of the perturbative BS approach to
$\pi^+\pi^-$ atom observables.
The existence of a quasistable $\pi^+\pi^-$ atom ($\tau\sim 10^{-15}~sec$)
corresponds to the
bound state pole in the four-point Green  function for the
transition $\pi^+\pi^-\rightarrow\pi^+\pi^-$ at a complex value of
the c.m. energy $P^2=\bar M^2=M^2-iM\Gamma$. Here $M$ denotes
the "mass" of an atom, and $\Gamma$ stands for the decay width.
Hereafter all formulae are restricted to the c.m.s. of an atom.

The BS w.f. of an atom $\chi_{BS}$ for $P^2\rightarrow\bar M^2$ obeys
the exact BS equation (Fig. 1a)
\eq\label{exact}
G_2^{-1}(P;p)\chi_{BS}(P;p)=\int\frac{d^4q}{(2\pi)^4}V_{BS}(P;p,q)\chi_{BS}(P;q)
\en
\noindent
Here $G_2(P;p)=D(\frac{1}{2}P+p)D(\frac{1}{2}P-p)$ is the two-pion Green
function where
$D(k)$ stands for the dressed pion propagator. Further, $V_{BS}$ denotes
the BS equation kernel, which is a sum of all four-point one-particle
irreducible diagrams with amputated external legs.

It is appropriate to "transfer" the self-energy insertions in the charged
pion external legs to the r.h.s of the BS equation. This can be easily achieved
if one defines $\chi(P;p)=G_0(P;p)G_2^{-1}(P;p)\chi_{BS}(P;p)$ and
$V(P;p,q)=V_{BS}(P;p,q)G_2(P;q)G_0^{-1}(P;q)$, where
$G_0(P;p)=i\bigl[\bigl(\frac{1}{2}P+p\bigr)^2-m_\pi^2\bigr]^{-1}\times
i\bigl[\bigl(\frac{1}{2}P-p\bigr)^2-m_\pi^2\bigr]^{-1}$ is the free
two-particle Green  function and $m_\pi=m_{\pi^\pm}$ denotes the
charged pion mass. The diagrammatic expansion of the new kernel $V$
is given in Fig. 1b. It, in addition to the diagrams included in the
"true" kernel $V_{BS}$, contains the self-energy diagrams in
{\it outgoing} external pion legs, i.e., only {\it half} of the
possible insertions in external legs. Note that this property of the
new kernel $V$ is crucial for proving the gauge invariance of
bound-state characteristics, as well as for demonstrating
the cancellation of infrared singularities in these characteristics
(see below). The BS equation for the new w.f. $\chi$ depicted in
Fig. 1c is given by
\eq\label{fullBS}
G_0^{-1}(P;p)\chi(P;p)=\int\frac{d^4q}{(2\pi)^4}V(P;p,q)\chi(P;q)
\en
The kernel $V$ contains the instantaneous Coulomb part $V_C$ which, in a
complete analogy with the positronium case, is responsible for the formation
of the bound state composed of $\pi^+$ and $\pi^-$. We are willing to
develop the perturbative expansion of the atom observables in the
"remainder" of the potential denoted by $V'=V-V_C$. For this purpose
we give first a complete solution of the "unperturbed" problem, with the
kernel containing only the instantaneous Coulomb part.

We choose the instantaneous Coulomb part of the potential according
to the Barbieri-Remiddi prescription \cite{Barbieri}, to be \cite{JETP}
\eq\label{BARBIERI}
V_C(\vec p, \vec q\,)=(w(\vec p\,))^{1/2}\,\,
\frac{4im_\pi e^2}{(\vec p-\vec q\,)^2}\,\,(w(\vec q\,))^{1/2},
\quad\quad
w(\vec p\,)=\bigl( m_\pi^2+\vec p^{~2}\bigr)^{1/2}
\en
Note that a particular choice of the Barbieri-Remiddi kernel (\ref{BARBIERI})
is only the matter of convenience and the final results are not affected
by this choice (below we shall demonstrate this property of the perturbative
expansion explicitly). However, choosing (\ref{BARBIERI}) as an initial
kernel, one can take advantage of the fact that the BS equation with
a kernel of this sort is exactly solvable, with the properly normalized
ground-state solution given by \cite{JETP}
\eq\label{COULOMBWF}
\psi_C(p)=iG_0(\Ms;p)\,\,4\,(w(\vec p\,))^{1/2}\,\,
\frac{4\pi\alpha m_\pi\phi_0}{\vec p^{~2}+\gamma^2},
\quad\quad
\bar\psi_C(p)=\psi_C(p)
\en
\noindent
where $\gamma=\frac{1}{2}m_\pi\alpha$, $\pi\phi_0^2=\gamma^3$ and
$(\Ms)^2=m_\pi^2(4-\alpha^2)$ is the eigenvalue corresponding to the
unperturbed ground-state solution. The c.m.s. momentum in the free
Green function $G_0$ in Eq. (\ref{COULOMBWF}) has the components
$(\Ms,\vec 0\,)$.

The normalization condition for the Coulomb w.f. reads as
\eq\label{NORM}
<\psi_C|N(\Ms)|\psi_C>=1,
\quad\quad
N(\Ms;p,q)=(2\pi)^4\delta^{(4)}(p-q)\,\,
\frac{i}{2\Ms}\,\frac{\partial}{\partial\Ms}\,G_0^{-1}(\Ms;p)
\en
and the scalar product in the momentum space is defined by the integral
over $d^4q/(2\pi)^4$. We shall use this shorthand notation hereafter.

The exact solution for the Green function corresponding to the
nonrelativistic Coulomb problem, was given by Schwinger \cite{Schwinger}.
Using this result, one can obtain the solution for the 4D Coulomb
Green  function corresponding to the kernel (\ref{BARBIERI}) \cite{JETP}
\eq
G_C(\Ps;p,q)=(2\pi)^4\delta^{(4)}(p-q)G_0(\Ps;p)+
G_0(\Ps;p) T_C(\Es;\vec p,\vec q\,)G_0(\Ps;q)
\en
Where
\eq\label{SCHWINGER}
T_C(\Es;\vec p;\vec q\,)=16i\pi m_\pi\alpha\biggl[
\frac{1}{(\vec p-\vec q\,)^2}+
\int\limits_0^1\frac{\nu d\rho\rho^{-\nu}}{D(\rho;\Es;\vec p,\vec q\,)}\biggr]\\[2mm]
D(\rho;\Es;\vec p,\vec q\,)=(\vec p-\vec q\,)^2-
\frac{m_\pi}{4\Es}\biggl(\Es-\frac{\vec p^{~2}}{m_\pi}\biggr)
\biggl(\Es-\frac{\vec q^{~2}}{m_\pi}\biggr)(1-\rho)^2\nonumber\\[2mm]
\nu=\alpha\biggl(\frac{m_\pi}{-4\Es}\biggr)^{1/2}\quad\quad
\Es=\frac{(\Ps)^2-4m_\pi^2}{4m_\pi}\nonumber
\en
The first and second terms in the r.h.s. of Eq. (\ref{SCHWINGER})
correspond to the exchange of one and multiple Coulomb photons, respectively.
In the vicinity of the bound-state pole $(\Ps)^2\rightarrow(\Ms)^2$,
$\nu\rightarrow 1$ and the integral in the r.h.s. of Eq. (\ref{SCHWINGER})
diverges as $\int_0^1d\rho/\rho$. Extracting this divergent piece which
corresponds to the bound-state pole in the Green function, one can write
\eq
G_C(\Ps;p,q)=i\,\frac{\psi_C(\Ms;p)\bar\psi_C(\Ms;q)}{(\Ps)^2-(\Ms)^2+i0}+
G_R(\Ps;p,q)
\en
where in the vicinity of the bound-state pole the regular part of the
Coulomb Green  function takes the form
\eq\label{REGULAR}
G_R(\Ms;p,q)&=&
(2\pi)^4\delta^{(4)}(p-q)G_0(\Ms;p)\nonumber\\[2mm]
&+&i\bigr( w(\vec p\,)w(\vec q\,)\bigl)^{1/2}
\biggl[\tilde\Phi(\vec p,\vec q\,)-
S(\vec p\,)S(\vec q\,)\frac{8}{\Ms}\frac{\partial}{\partial\Ms}\biggr]
G_0(\Ms;p)G_0(\Ms;q)
\nonumber\\[2mm]
\tilde\Phi(\vec p,\vec q\,)&=&16\pi m_\pi\alpha\biggl[
\frac{1}{(\vec p-\vec q\,)^2}+I_R(\vec p,\vec q\,)\biggr]
+\frac{1}{(m_\pi\alpha)^2}S(\vec p\,)S(\vec q\,)\tilde R(\vec p,\vec q\,)
\\[2mm]
S(\vec p\,)&=&\frac{4\pi m_\pi\alpha\phi_0}{\vec p^{~2}+\gamma^2},
\quad\quad
\tilde R(\vec p,\vec q\,)=20-\biggl(\frac{8}{\pi m_\pi\alpha}\biggr)^{1/2}
\bigl[S(\vec p\,)+S(\vec q\,)\bigr]
\nonumber\\[2mm]
I_R(\vec p,\vec q\,)&=&\int_0^1\frac{d\rho}{\rho}
\bigl[ D^{-1}(\rho;-\frac{1}{4}m_\pi\alpha^2;\vec p,\vec q\,)-
D^{-1}(0;-\frac{1}{4}m_\pi\alpha^2;\vec p,\vec q\,)\bigr]\nonumber
\en

The solution $\chi(P;p)$ of the exact BS equation (\ref{fullBS}) can
be expressed via the unperturbed solution $\psi_C(\Ms;p)$ by the
following limiting procedure \cite{PLB,JETP}
\eq\label{CONNECTION}
<\chi|=C<\psi_C|G_C^{-1}(\Ps)G(P),
\quad\quad
(\Ps)^2\rightarrow (\Ms)^2,\quad
P^2\rightarrow\bar M^2
\en
where $C$ denotes the normalization constant. Note that this relation
is the relativistic generalization of the well-known nonrelativistic formula
\eq
<\chi|=\lim_{\eta\to 0(+)}\,\,i\eta\,\,<\psi_0|\,\frac{1}{E-H+i\eta}
\en
which connects eigenvectors of the total Hamiltonian $H$ with
the unperturbed eigenvectors (see, e.g., \cite{Goldberger}).

The result in Eq. (\ref{CONNECTION}) depends on details of the
limiting procedure. This equation makes sense if the quantities
$(\Ps)^2-(\Ms)^2$ and $P^2-\bar M^2$ are assumed to be the infinitesimal
variables of equal strength. In Refs. \cite{PLB,JETP} we have assumed
the prescription
$(\Ps)^2=(\Ms)^2+\lambda,~P^2=\bar M^2+\lambda,~\lambda\rightarrow 0$.
Note that we can employ this prescription without the loss in generality,
since the change of the direction in the $((\Ps)^2,P^2)$ plane
along which this limiting procedure is performed affects only
the normalization constant $C$. Further, the validity of Eq. (\ref{CONNECTION})
can be trivially checked by extracting the bound-state pole in $G(P)$
and using the BS equation for $<\psi_C|$.

Let us now introduce the relativistic generalization of the projector
operator onto the states orthogonal to the ground-state solution
\eq\label{PROJECTOR}
Q=1\,-\, N(\Ms)|\psi_C><\psi_C|
\en
Then, with the use of the Hilbert identity it is easy to demonstrate
that Eq. (\ref{CONNECTION}) can be rewritten as follows
\eq\label{CHI}
<\chi|&=&<\chi|N(\Ms)|\psi_C><\psi_C|\biggl(
1\,+\, V'(P)\bigl[G_0^{-1}(\Ps)-V_C-QV'(P)\bigr]^{-1}Q\biggr)
\nonumber\\[2mm]
&-&<\chi|\Delta G_0^{-1}\bigl[G_0^{-1}(\Ps)-V_C-QV'(P)\bigr]^{-1}Q
\en
where $\Delta G_0^{-1}=G_0^{-1}(P)-G_0^{-1}(\Ps)$ and the limiting procedure
is implicit. In the derivation of Eq. (\ref{CHI}) we have used
\eq
<\psi_C|G_C^{-1}(\Ps)\bigl[G_0^{-1}(\Ps)-V_C-QV'(P)\bigr]^{-1}=0
\en
which stems from the fact that the inverse operator in the l.h.s.
of this equation does not have a pole in this limit.
With the limiting prescription chosen above the normalization constant
equals \cite{PLB}
\eq
-C^{-1}=<\chi|N(\Ms)|\psi_C>
\en

Equation (\ref{CHI}) can be solved with respect to $<\chi|$, resulting in
\eq\label{PRELIMINARY}
<\chi|=-C^{-1}<\psi_C|
\,\,\bigl[ 1+V'(P)G_VQ\bigr]\,\,\bigl[ 1+\Delta G_0^{-1}G_VQ\bigr]^{-1}
\en
where the operator $G_VQ$ obeys the equation
\eq
G_VQ=G_R(\Ps)Q+G_R(\Ps)QV'(P)G_VQ
\en
and $G_R$ stands for the regular part of the Coulomb Green function.
It can be easily demonstrated that Eq. (\ref{PRELIMINARY}) can be
rewritten as
\eq
<\chi|=-C^{-1}<\psi_C|\bigl[1+(\Delta G_0^{-1}-V'(P))G_RQ\bigr]^{-1}
\en
Substituting this solution into the complete BS equation (\ref{fullBS})
and using the BS equation for the function $|\psi_C>$, we arrive at the final
relation
\eq\label{BASIC}
<\psi_C|\bigl[ 1+(\Delta G_0^{-1}-V'(P)) G_R Q\bigr]^{-1}
(\Delta G_0^{-1}-V'(P))|\psi_C>=0
\en
which provides the basis for the perturbative expansion of bound-state
observables (the energy of the atomic level and decay width). Namely,
the only unknown quantity in the l.h.s. of Eq. (\ref{BASIC}) is the
bound-state total four-momentum $P$, which enters parametrically into this
expression. Expanding the l.h.s. of Eq. (\ref{BASIC}) in the perturbation
theory up to a given order, one can then determine bound-state
observables with a required accuracy. Note also that Eq. (\ref{BASIC})
is a complex equation, and in every perturbative order
it provides two real equations for determining the energy level shift
and decay width.

Equation (\ref{BASIC}), however, still contains the BS kernel, and does not
contain the $\pi\pi$ scattering amplitudes. Below we carry out the first-order
perturbative calculations and demonstrate explicitly that only
these amplitudes appear in the final result. For this purpose let us
note first that the quantity $G_RQ$ in Eq. (\ref{BASIC})
is given by formulae similar to (\ref{REGULAR}), with
$\tilde\Phi$ and $\tilde R$ replaced by $\Phi$ and $R$, respectively, and
\eq\label{TILDE}
R(\vec p,\vec q\,)=\tilde R(\vec p,\vec q\,)\,\,+\,\, 5+ \,\,
{\mbox{higher orders in }} \alpha
\en
(this can be demonstrated by straightforward calculations,
using Eqs. (\ref{COULOMBWF}), (\ref{NORM}), (\ref{REGULAR}) and (\ref{PROJECTOR})).
We isolate the free part in $G_RQ$ by writing $G_RQ=G_0(\Ms)+\delta G$.

Let us now turn to the perturbation kernel $V'(P)$.
This potential can be decomposed into the following parts:

\noindent
1. A purely strong part, which is isotopically invariant. This part
survives when electromagnetic interactions are "turned off" the theory.

\noindent
2. The part which is responsible for the $m_{\pi^\pm}-m_{\pi^0}$
electromagnetic mass difference

\noindent
3. Remaining electromagnetic effects, including the exchanges of
virtual photons.

Parts 1 and 2 are regarded to be more important for the following
reasons. The first term includes strong interactions which govern the
decay of a pionium. The second term makes this decay kinematically allowed.
Consequently, it seems to be natural to consider them together,
denoting the corresponding potential as $V_{12}=V_1+V_2$. The $T$-matrix
corresponding to summation of the potential $V_{12}$ in all orders
is given by $T_{12}(P)=V_{12}(P)+V_{12}(P)G_0(P)T_{12}(P)$. The rest
of the potential is referred to as $V_3=V'-V_{12}$ and is treated
perturbatively.

We would like to emphasize once more that this splitting is rather
convention-dependent and is dictated by the convenience considerations.
In practice it is convenient to include, into parts 1 and 2, as much
terms as possible. It is obvious, however, that the final results
do not depend on the prescription chosen for that splitting.

We perform the perturbative expansion of the basic equation (\ref{BASIC})
in $V_3$ and $\delta G$ up to the first nontrivial order. Meanwhile
we expand $\Delta G_0^{-1}$ in the Taylor series in the variable
$\delta M=\bar M-\Ms$ and substitute
\eq\label{MASS}
\bar M=\Ms+\Delta E^{(1)}+\Delta E^{(2)}-\frac{i}{2}\Gamma^{(1)}
-\frac{i}{2}\Gamma^{(2)}+\frac{1}{8\Ms}(\Gamma^{(1)})^2+\cdots
\en
Expressing everywhere $V_{12}$ in terms of $T_{12}$, we finally arrive
at the identity
\eq\label{LARGE}
0&=&-2i\Ms\delta M-<\psi_C|T_{12}|\psi_C>\nonumber\\[2mm]
&+&\delta M<\psi_C|(G_0^{-1})'G_0T_{12}|\psi_C>
+\frac{1}{2}(\delta M)^2<\psi_C|(1+T_{12}G_0)(G_0^{-1})''
+(G_0^{-1})''G_0T_{12}|\psi_C>
\nonumber\\[2mm]
&+&<\psi_C|(\delta M(G_0^{-1})'-T_{12})\delta GT_{12}|\psi_C>
-<\psi_C|(1+T_{12}G_0)V_3(1+G_0T_{12})|\psi_C>
\en
where $G_0=G_0(\Ms)$ and the prime stands for the differentiation with
respect to $\Ms$.

Equation (\ref{LARGE}) contains all first-order corrections
to the pionium lifetime. Below we examine this relation term
by term. Note that we employ the commonly accepted "local"
approximation, i.e. we assume that the quantity $T_{12}$ does not depend
on relative momenta. The origin of this approximation can be traced
to the "sharpness" of the Coulomb w.f. of an atom which in the
momentum space has the characteristic range $\gamma\sim 1~MeV$, much
smaller than the typical hadronic scales.

\section{Relativistic Deser-type formulae with lowest-order corrections}

In the lowest-order calculations only the first two terms in the r.h.s.
of Eq. (\ref{LARGE}) contribute. Further,
one can assume $\psi_C(0)=\int d^4k/(2\pi)^4\,\psi_C(\Ms;k)=m_\pi^{-1/2}\phi_0$
in this approximation. Then, taking the real and imaginary part of
Eq. (\ref{LARGE}), we arrive at
\eq\label{rel-deser}
\Delta E^{(1)}={\mbox{Re}}\biggl( \frac{iT_{12}}{2\Ms m_\pi}\phi_0^2\biggr),
\quad\quad
-\frac{1}{2}\Gamma^{(1)}={\mbox{Im}}\biggl(\frac{iT_{12}}{2\Ms m_\pi}\phi_0^2\biggr)
\en
Further, we can write
\eq\label{UNITARITY}
{\mbox{Re}}\,(iT_{12})&=&
16\pi\,{\cal T}_{\pi^+\pi^-\rightarrow\pi^+\pi^-}(4m_\pi^2;\vec 0,\vec 0\,)
\nonumber\\[2mm]
{\mbox{Im}}\,(iT_{12})&=&
-16\pi\,\biggl(\frac{\Delta m_\pi}{2m_\pi}\biggr)^{1/2}
\biggl( 1-\frac{\Delta m_\pi}{2m_\pi}\biggr)^{1/2}
\,\,|{\cal T}_{\pi^+\pi^-\rightarrow\pi^0\pi^0}(4m_\pi^2;\vec 0,\vec q_0\,)|^2,\quad
\en
where ${\cal T}(s;\vec p,\vec q\,)$ denote the (dimensionless) $S$-wave
$\pi\pi$ scattering amplitudes and $\vec q_0$ is the relative momentum
of the $\pi^0\pi^0$ pair at the threshold $s=4m_\pi^2$, with the magnitude
given by the relation $m_\pi^2=m_{\pi^0}^2+\vec q_0^{~2}$.

We would like to emphasize that the second relation in Eq. (\ref{UNITARITY})
differs from an analogous relation given in Ref. \cite{Sazdjian},
though the starting equations (\ref{rel-deser}) in the both papers coincide.
In Ref. \cite{Sazdjian} the magnitude of the three-momentum both for
$\pi^+\pi^-$ and $\pi^0\pi^0$ pairs was set equal to $0$. Consequently,
neutral pions in the final state turned out to be off shell. Contrary
to Ref. \cite{Sazdjian} we deduce from the Cutcosky rule that
neutral pions in Eq. (\ref{UNITARITY}) are on-shell.
Note that this discrepancy with Ref. \cite{Sazdjian} leads to
{\it different} predictions for the corrections e.g., due to the
mass difference $m_{\pi^\pm}-m_{\pi^0}$ in one loop (see below).

Substituting Eq. (\ref{UNITARITY}) into (\ref{rel-deser}), we reproduce
the lowest-order relativistic Deser-type formulae for the energy level
shift and lifetime of the pionium
\eq\label{calT}
&&\Delta E=\frac{4\pi}{m_\pi^2}\,\,\,
{\cal T}_{\pi^+\pi^-\rightarrow\pi^+\pi^-}\,\,\,\phi_0^2
\nonumber\\[2mm]
&&\Gamma=\frac{16\pi}{m_\pi^2}\,
\biggl(\frac{2\Delta m_\pi}{m_\pi}\biggr)^{1/2}
\biggl( 1-\frac{\Delta m_\pi}{2m_\pi}\biggr)^{1/2}
\,\,|{\cal T}_{\pi^+\pi^-\rightarrow\pi^0\pi^0}|^2\,\,\phi_0^2
\en

\subsection{Relativistic correction to the pionium w.f.}

The correction in the pionium lifetime coming from this effect is contained
in the second term of Eq. (\ref{LARGE}). Namely, up to $O(\alpha)$ terms
\eq\label{c0}
\psi_C(0)=\int\frac{d^4k}{(2\pi)^4}\,\,\psi_C(\Ms;k)=\frac{\phi_0}{m_\pi^{1/2}}
(1-C_0\alpha), \quad\quad
C_0=0.381\cdots
\en
(details can be found in Appendix A). Thus, up to order $O(\alpha)$ in the
local approximation the second term in Eq. (\ref{LARGE}) reads as
$m_\pi^{-1}\,\,T_{12}\,\,\phi_0^2\,\,(1-2C_0\alpha)$. The term proportional
to $C_0\alpha$ in this expression induces the corresponding correction
in the pionuim lifetime. Note that the value of this correction
is determined by the expression of the unperturbed solution $\psi_C$
(\ref{COULOMBWF}) and hence depends on the particular choice of the
instantaneous Coulomb part of the potential (for our case the Barbieri-Remiddi
prescription (\ref{BARBIERI})). We shall see, however, that in the
full expression for the correction to the atom lifetime the term proportional
to $C_0\alpha$ disappears, indicating that the final results do not depend
on a particular choice of the zeroth-order kernel.

\subsection{Correction due to the displacement of the bound-state pole
by strong interactions}

This correction is induced by third and fourth terms in Eq. (\ref{LARGE}).
The calculation of this sort of integrals is carried out in
a straightforward way was considered in Ref. \cite{PLB}. Below we give
the result of these calculations

\eq
<\psi_C|(G_0^{-1})'G_0T_{12}|\psi_C>&=&
T_{12}\,\,\psi_C(0)\,\,
\int\frac{d^4p}{(2\pi)^4}\bar\psi_C(\Ms;p)(G_0^{-1}(\Ms;p))'G_0(\Ms;p)
\nonumber\\[2mm]
\label{A}
&=&\frac{i}{\alpha^2}\,\,\frac{iT_{12}\phi_0^2}{m_\pi^2}+\cdots
\\[2mm]
\label{B}
<\psi_C|(G_0^{-1})''|\psi_C>&=&
\int\frac{d^4p}{(2\pi)^4}\bar\psi_C(\Ms;p)(G_0^{-1}(\Ms;p))''\psi_C(\Ms;p)
=\frac{10i}{\alpha^2}+\cdots
\\[2mm]
<\psi_C|T_{12}G_0(G_0^{-1})''|\psi_C>&=&
\psi_C(0)\,\, T_{12}\,\,
\int\frac{d^4p}{(2\pi)^4}G_0(\Ms;p)(G_0^{-1}(\Ms;p))''\psi_C(\Ms;p)
\nonumber\\[2mm]
\label{C}
&=&-\frac{2i}{\alpha^2}\,\,\frac{iT_{12}\phi_0^2}{m_\pi^3}+\cdots
=\, <\psi_C|(G_0^{-1})''G_0T_{12}|\psi_C>
\en
and ellipses stand for higher-order terms in $\alpha$.
From Eq. (\ref{rel-deser}) it is easy to see that the real and
imaginary parts of integral
(\ref{C}) are down by small factors $\Delta E^{(1)}/m_\pi$ and
$\Gamma^{(1)}/m_\pi$ as compared to integral (\ref{B}). Therefore
further we shall neglect (\ref{C}) in our calculations.

Note that Eq. (\ref{B}) contains the second derivative of the inverse
free Green  function $G_0$ with respect to the bound-state mass.
Hence this is a true relativistic correction arising from the
BS treatment of the bound-state problem since in the nonrelativistic
case the free inverse Green  function is linear in the bound-state
energy (cf. Ref. \cite{PLB})

\subsection{Correction due to the exchange of Coulomb photons}

This correction stems from the fifth term of Eq. (\ref{LARGE}).
The calculation of the corresponding integrals is considered in Appendix A.
Below we give the results
\eq
<\psi_C|(G_0^{-1})'\delta GT_{12}|\psi_C>&=&
T_{12}\,\,\psi_C(0)\,\,
\int\frac{d^4p}{(2\pi)^4}\frac{d^4q}{(2\pi)^4}\bar\psi_C(\Ms;p)
(G_0^{-1}(\Ms;p))'\delta G(p,q)
\nonumber\\[2mm]\label{AA}
&=&-\frac{i}{\alpha^2}\frac{iT_{12}\phi_0^2}{m_\pi^2}+\cdots
\\[2mm]
<\psi_C|T_{12}\delta GT_{12}|\psi_C>&=&
(T_{12})^2\,\,(\psi_C(0))^2\,\,
\int\frac{d^4p}{(2\pi)^4}\frac{d^4q}{(2\pi)^4}\delta G(p,q)
\nonumber\\[2mm]\label{BB}
&=&
\frac{i\alpha}{16\pi m_\pi}\,\,({\rm ln}\alpha-2.694)
(T_{12})^2\,\,\phi_0^2+\cdots
\en
where the term nonanalytic in the fine structure constant (containing
${\rm ln}\alpha$ in Eq. (\ref{BB})) comes from the infrared-singular
one-photon exchange piece in the Coulomb Green function (Eqs.
(\ref{REGULAR}), (\ref{TILDE})).

With the calculated integrals and the lowest-order
relations (\ref{rel-deser}), it is a simple algebraic task to derive, from
Eqs. (\ref{MASS}) and (\ref{LARGE}), the first-order
correction to the pionium decay width
\cite{JETP}
\eq\label{fn}
\hspace*{-.7cm}\Gamma^{(2)}&=&\Gamma^{(1)}\biggl\{
\underbrace{\biggl(-\frac{9}{8}\,\frac{\Delta E^{(1)}}{E_1}\biggr)}_{\rm strong}\,\,\,
+\underbrace{(-2C_0\alpha)}_{\rm relativistic~w.f.}\, + \,\,\,\,
\underbrace{\left({1}/{2}+2.694-{\rm ln}\alpha\right)
\frac{\Delta E^{(1)}}{E_1}}_{\rm Coulomb~photon~exchanges}\,\,\,
\nonumber\\
\hspace*{-.7cm}&-&{\left( M^\star\Gamma^{(1)}\right)^{-1}}
{\rm Re}<\psi_C|(1+T_{12}G_0)V_3(1+G_0T_{12})|\psi_C>\biggr\}
\en
where $E_1$ stands for the energy of the unperturbed ground-state
level $E_1=-1/4\,\,m_\pi\alpha^2$.

Comparing the Eq. (\ref{fn}) with the corresponding expression given
in Ref. \cite{Sazdjian} (referred to as "second-order strong correction"),
it is easy to see that our expression for the
contribution of Coulomb photon exchanges contains an additional
${\rm ln}\alpha$ term. The origin for this disagreement can be
easily established. Namely, the authors of Ref. \cite{Sazdjian}
include an additional contribution from the so-called "constraint diagram"
into this term. This diagram cancels explicitly the one-Coulomb photon
exchange term, leaving only multiphoton exchanges. We have checked that,
having merely discarded this term, after our calculations we come to the
result numerically very close to that given in Ref. \cite{Sazdjian}.
The true result, however, cannot depend on the formalism used for the
description of a bound state, either the BS equations or equations of the 3D
constraint theory; So, one can ask whether the contribution
regarded as the counterpart of the "constraint diagram" exists in the BS framework
for bound states. The answer is yes, this diagram is contained in the
"electromagnetic kernel" $V_3$ in Eq. (\ref{fn}). The reason why
we include this diagram into $V_3$ rather than in the "second-order
correction" is simple: this diagram is accompanied by the diagram
of virtual photon exchange (see Fig. 2d)
neglected in Ref. \cite{Sazdjian}. The latter diagram
also produces the ${\rm ln}\alpha$ term which exactly cancels
the corresponding term from the "constraint diagram" (below we shall discuss
this in more detail). The remainder is regular in the fine structure
constant. Consequently, we find more natural at the present
stage to omit both these contributions on the equal footing, rather than
to retain only one of them, namely the "constraint diagram".

We would like to mention here that the sign and magnitude of the
nonanalytic term appears to be exactly the same as in the nonrelativistic
treatment of the pionium, indicating that, as one could expect from the
beginning, the "electromagnetic kernel" produces the corrections which
are analytic in $\alpha$ (at least in the lowest order). In the scattering
theory, when the electromagnetic corrections are taken into account,
the expression for the decay width corresponding to the second relation
from Eq. (\ref{rel-deser}) ${\rm Im} iT_{12}$ is replaced by ${\rm Im} iT_{cc}$,
where $T_{cc}$ denotes the scattering amplitude of charged particles
at threshold in the presence of the Coulomb potential \cite{Rasche1,Rasche2}. Accordingly, it
leads to the replacement of ${\rm Im}\, a_{had}$ by ${\rm Im}\, a_{cc}$,
where $a_{had}$ and $a_{cc}$ denote the "hadronic" and exact scattering
lengths of charged particles. However, assuming that the hadronic
potential has a finite range denoted by $R$, the following relation
between $a_{had}$ and $a_{cc}$ can be established \cite{Trueman,Goldberger}
\eq
\frac{1}{a_{cc}}=\frac{1}{a_{had}}
-\frac{2}{r_B}{\rm ln}\biggl(\frac{2R}{r_B}\biggr)
+\frac{1}{r_B}\biggl({\mbox{series in powers of }}\frac{2R}{r_B}\biggr)
\en
where $r_B$ is the Bohr radius of the pionium which is inverse proportional
to $\alpha$. So, up to the logarithmic terms,
\eq
{\rm Im}\, a_{cc}={\rm Im}\, a_{had}
\biggl( 1-\frac{4{\rm Re}\, a_{had}}{r_B}\,{\rm ln}\alpha\biggr)=
{\rm Im}\, a_{had}\biggl( 1-\frac{\Delta E^{(1)}}{E_1}{\rm ln}\alpha\biggr)
\en
where in the second relation we have used the Deser formula for the
atom energy level displacement. Consequently, the decay width is modified
according to $\Gamma\rightarrow\Gamma ( 1-\Delta E^{(1)}{\rm ln}\alpha /E_1 )$
(cf. Eq. (\ref{fn})).

In addition, we would like to note that the logarithmic term which,
just as in our approach, is of the second order in strong interactions
($\sim (T_{12})^2$) was found by Roig and Swift \cite{Roig}.
The authors of Ref. \cite{Roig} have studied the electromagnetic radiative
corrections to the $\pi\pi$ scattering and discovered the term
proportional to ${\rm ln}p$ in the amplitude. When substituted into bound-state
integrals, after rescaling, as usual, the integration momenta by
$p\rightarrow \gamma p$, this term produces the ${\rm ln}\alpha$ correction,
as in Eq. (\ref{fn}).

So far our treatment of the pionium decay width has been incomplete.
Now we turn to the calculation of corrections induced by the
last term in Eq. (\ref{LARGE}) containing the "electromagnetic" kernel
$V_3$.

\subsection{Mass shift and radiative corrections}

Into the kernel $V_3$ we include the following diagrams: the diagram with
the "residual" photon exchange (i.e. the virtual photon exchange minus
Coulomb potential), Fig. 2, the self-energy corrections to outgoing
pion legs, Fig. 3, the vacuum polarization diagram, Fig. 4 and
the vertex corrections, Fig. 5.
The contributions containing low-energy constants and
tadpole terms are included into $T_{12}$. Thus, we take here the advantage
of the arbitrariness in splitting the potential and include all terms into
$T_{12}$ which are {\it a priori} known to have a smooth momentum dependence
on the bound-state scale $\gamma$. Only the potentially "dangerous"
terms which contain the photon propagator with vanishing mass are
to be treated with the bound-state equation.

In this section we are concerned with the first two terms in $V_3$. They
read as

1. Residual photon exchange (Fig. 2)
\eq
V_\gamma-V_C=ie^2(P+p+q)_\mu(P-p-q)_\nu D^{\mu\nu}(p-q)-V_C(\vec p,\vec q\,)
\en
where $D^{\mu\nu}$ denotes the photon propagator. The calculations
are most easily carried in the Coulomb gauge with
\eq
D^{00}(\vec k\,)=-\frac{1}{\vec k^{~2}},
\quad
D^{0i}(\vec k\,)=D^{i0}(\vec k\,)=0,
\quad
D^{ij}(\vec k\,)=-\biggl(\delta^{ij}-\frac{k^ik^j}{\vec k^{~2}}\biggr)
\frac{1}{k^2+i0}
\en
The corresponding matrix element equals
\eq
{\rm Re}
<\psi_C|(V_\gamma-V_C)+T_{12}G_0(V_\gamma-V_C)+(V_\gamma-V_C)G_0T_{12}+
T_{12}G_0(V_\gamma-V_C)G_0T_{12}|\psi_C>
\nonumber
\en
It is easy to observe that ${\rm Re}<\psi_C|(V_\gamma-V_C)|\psi_C>=0$ at the
bound state energy. Consequently, the first term in the matrix element
vanishes. The following two terms are equal for symmetry considerations.
Further, we completely neglect the fourth term in this matrix element
(Fig. 2d). This term, as we have mentioned before, contains the contribution
of "constraint diagram" given by $V_C$. However, it is obvious that
the $D^{00}$ component in the expression of $V_\gamma$ which has exactly
the same infrared singular behavior as the Coulomb potential
leads to the same ${\rm ln}\alpha$ nonanalytic term in the lifetime.
In the expression of $V_3$ these nonanalytic terms cancel and the remainder
is analytic in $\alpha$ (at least in the lowest order). For this reason
we find more safe to neglect the combination $V_\gamma-V_C$ rather than
$V_\gamma$ alone, as in Ref. \cite{Sazdjian}.

The remaining term in the expression of the matrix element is exactly of
the form of the "retardation correction" discussed in Ref. \cite{Silagadze}.
To demonstrate this, we note that
$|\delta\psi_\gamma>=G_0(V_\gamma-V_C)|\psi_C>$ gives the first-order
perturbative correction to the bound-state w.f. due to the retardation
effect (i.e. the difference between $V_\gamma$ and $V_C$). The correction
induced in the pionium lifetime due to this effect is given by
\eq\label{SILAGADZE}
\Gamma\rightarrow\Gamma\,\,\biggl( 1-\frac{1}{\Ms\Gamma^{(1)}}\,\,{\rm Re}[
2\,\psi_C(0)\, T_{12}\,\delta\psi_\gamma(0)]\,\,\biggr)=
\Gamma\,\,\biggl(1+\frac{2\delta\psi_\gamma(0)}{\psi_C(0)}\biggr)
\en
where the second relation was derived with the use of Eq. (\ref{rel-deser}).
From Eq. (\ref{SILAGADZE}) it is apparent that the kernel
$V_\gamma-V_C$ is responsible for the retardation correction in the
pionium lifetime.

Finally, the matrix element we are looking for can be written in the
following form
\eq
{\cal M}_\gamma=
{\rm Re}\biggl[\,2\times\psi_C(0)\times T_{12}\times
\int\frac{d^4q}{(2\pi)^4}\bigl(\Lambda_\gamma-\Lambda_C\bigr)
\psi_C(\Ms;q)\biggr]
\en
where
\eq
\Lambda_\gamma&=&-ie^2\int\frac{d^4p}{(2\pi)^4}\,\, G_0(\Ms;p)\,\,
(P+p+q)_\mu(P-p-q)_\nu \,\, D^{\mu\nu}(p-q)
\nonumber\\[2mm]
\Lambda_C&=&-\int\frac{d^4p}{(2\pi)^4}\,\, G_0(\Ms;p)\,\, V_C(\vec p,\vec q\,)
\en

2. The kernel corresponding to the insertion of self-energy graphs into the
outgoing pion legs (Fig. 3)
\eq
V_\Sigma=e^2\,\, V_{12}(P;p,q)\,\,\biggl(
\frac{\tilde\Pi\bigl(\frac{P}{2}+q\bigr)}{\bigl(\frac{P}{2}+q\bigr)^2-m_\pi^2}+
\frac{\tilde\Pi\bigl(\frac{P}{2}-q\bigr)}{\bigl(\frac{P}{2}-q\bigr)^2-m_\pi^2}
\biggr)
\en
where
\eq\label{PROPAGATOR}
\Pi(l)&=&i\int\frac{d^4k}{(2\pi)^4}\,\,\frac{1}{(l-k)^2-m_\pi^2}\,\,
(2l-k)_\mu (2l-k)_\nu\,\, D^{\mu\nu}(k)\nonumber\\[2mm]
\tilde\Pi(l)&=&\Pi(l)-\Pi(l^2=m_\pi^2),\quad\quad
Z(l)=\frac{\tilde\Pi(l)}{l^2-m_\pi^2},\quad\quad
Z_\pm(l)=Z\bigl(\frac{P}{2}\pm l\bigr)
\en
The corresponding matrix element can be rewritten as
\eq\label{fig3}
{\cal M}_{\Sigma}=<\psi_C|\,\, e^2\,\,T_{12}\,\,(Z_++Z_-)\,\,(1+G_0T_{12})\,\,
|\psi_C>
\en
We neglect here the term which is of second order in the strong interaction
amplitude. Then, the sum of matrix elements reads as
\eq\label{onehalf}
{\cal M}_\gamma+{\cal M}_\Sigma=
{\rm Re}\biggl[\,2\,\,\psi_C(0)\,\, T_{12}\,\,
\int\frac{d^4q}{(2\pi)^4}\bigl(\Lambda_\gamma-\Lambda_C
-\frac{e^2}{2}Z_+(q)-\frac{e^2}{2}Z_-(q)
\bigr)\psi_C(\Ms;q)\biggr]
\en
We would like to emphasize here that the coefficient $\frac{1}{2}$
which emerges naturally in front of the self-energy term in the
r.h.s. of Eq. (\ref{onehalf}) ensures that the correction to the
decay width calculated from (\ref{onehalf}) is gauge invariant.
Moreover, it is well known that in this particular combination of
vertex and self-energy diagrams the "photon mass" disappears in the
calculated width without inclusion of the "soft photon emission" terms, which
seem to be rather awkward in the context of the bound-state problem.
In its turn, the origin of the coefficient $\frac{1}{2}$ can be traced
back to Eq. (\ref{fullBS}), where, as was mentioned before, the kernel
$V$ contains only a half of all possible self-energy graphs attached
to the external pion legs.
Note also than in the $S$-matrix elements the origin of emerging the
coefficient $\frac{1}{2}$ is quite different. In the latter case
one takes into account all self-energy graphs in the external
legs. However, restricting these matrix elements on mass shell one encounters
an expression of the type $0/0$ which can be tackled, introducing
an explicit "smearing function" in the initial Lagrangian. Passing then
to the limit when the "smearing function" tends to the unity, one
discovers the coefficient $\frac{1}{2}$ which multiplies the contribution from
the self-energy diagram (for a detailed discussion see e.g., \cite{Schweber}).
Consequently, though the coefficient $\frac{1}{2}$ emerges in the bound
state and the scattering problems from different sources, one ends up
with the same expression in these two cases, being gauge invariant
and infrared finite at threshold.

The integral in the r.h.s. of Eq. (\ref{onehalf}) can be easily evaluated
in the Coulomb gauge, bearing in mind that the final result is gauge
invariant.

In the calculation of the contribution from $\Lambda_\gamma$ one can use
the fact that the $D^{ij}$ component of the photon propagator in the Coulomb
gauge contributes only in order $O(\alpha^2{\rm ln}\alpha)$ in the decay
width and thus can be neglected. This considerably simplifies the
calculations. The result reads as (details can be found in Appendix B)
\eq\label{em-radiative}
&&\int\frac{d^4q}{(2\pi)^4}\Lambda_\gamma\psi_C(\Ms;q)\\[2mm]
&=&\int\frac{d^4p}{(2\pi)^4}\frac{d^4q}{(2\pi)^4}
G_0(\Ms;p)\frac{ie^2((\Ms)^2-(p_0+q_0)^2)}{(\vec p-\vec q\,)^2}
G_0(\Ms;q)\,\, 4i(w(\vec q\,))^{1/2}
\frac{4\pi\alpha m_\pi\phi_0}{\vec q^{~2}+\gamma^2}\nonumber\\[2mm]
&=&\frac{\phi_0}{m_\pi^{1/2}}\biggl(1+\frac{\alpha}{4\pi}\,N_\epsilon-
\frac{3\alpha}{2\pi}\biggr)+\cdots\nonumber
\en
where the dimensional regularization was used to handle the ultraviolet
divergences, and
\eq
N_\epsilon=\frac{1}{2-n/2}+\Gamma'(1)+{\rm ln}\, 4\pi-
{\rm ln}\biggl(\frac{m_\pi^2}{\mu^2}\biggr)
\nonumber
\en
with $n$ being the dimension of space and $\mu$ the mass scale used
in dimensional regularization.

The contribution containing $\Lambda_C$ can be trivially carried out
\eq
&&\int\frac{d^4q}{(2\pi)^4}\Lambda_C\psi_C(\Ms;q)\\[2mm]
&=&\int\frac{d^4p}{(2\pi)^4}\frac{d^4q}{(2\pi)^4}
G_0(\Ms;p)\frac{4im_\pi e^2(w(\vec p\,)w(\vec q\,))^{1/2}}{(\vec p-\vec q\,)^2}
G_0(\Ms;q)\,\, 4i(w(\vec q\,))^{1/2}
\frac{4\pi\alpha m_\pi\phi_0}{\vec q^{~2}+\gamma^2}\nonumber\\[2mm]
&=&m_\pi e^2\int\frac{d^3\vec p}{(2\pi)^2}\frac{d^3\vec q}{(2\pi)^2}
\frac{1}{w(\vec p\,)^{1/2}}\frac{1}{\vec p^{~2}+\gamma^2}
\frac{1}{(\vec p-\vec q\,)^2}
\frac{4\pi\alpha m_\pi\phi_0}{(\vec q^{~2}+\gamma^2)^2}=
\frac{\phi_0}{m_\pi^{1/2}}(1-C_0\alpha)+\cdots\nonumber
\en

The pion self-energy graph (\ref{PROPAGATOR}) calculated within the
dimensional regularization scheme in the Coulomb gauge is given by
\eq\label{PI}
\Pi(l)&=&
-\biggl(\frac{3m_\pi^2}{16\pi^2}N_\epsilon+\frac{7m_\pi^2}{16\pi^2}\biggr)
-\Delta\biggl(\frac{1}{8\pi^2}N_\epsilon+\frac{1}{4\pi^2}\biggr)
\\[2mm]
&+&\biggl[-\frac{\Delta}{8\pi^2}L-
\frac{\vec l^{2}}{6\pi^2}\frac{\Delta}{m_\pi^2+\Delta}
{\rm ln}\biggl(-\frac{\Delta}{m_\pi^2}\biggr)-
\frac{(\vec l^{~2})^2}{3\pi^2}\bigl( I(\vec l^{~2};\Delta)-I(\vec l^{~2};0)\bigr)
\biggr]\nonumber
\en
where $\Delta=l^2-m_\pi^2+i0$ and
\eq
L&=&2+\frac{w(\vec l\,)}{|\vec l\,|}{\rm ln}\,\biggl(
\frac{w(\vec l\,)-|\vec l\,|}{w(\vec l\,)+|\vec l\,|}\biggr),
\quad
L=O(\vec l^{~2}) {\mbox {   at small $|\vec l\, |$}}
\nonumber\\[2mm]
I(\vec l^{~2},\Delta)&=&\int_0^1 dx\int_0^1 du\,\,
\frac{xu^4}{x(1-u^2)\vec l^{~2}+xm_\pi^2-(1-x)\Delta}
\en
Note that the terms in square brackets in the r.h.s. of Eq. (\ref{PI})
are of higher order in $|\vec l\,|$ and/or $\Delta$ as compared to
the second term in the round brackets. Consequently, in the calculations
in the lowest order in $\alpha$ the term in square brackets can be
dropped, since it contributes only in order $\alpha^2{\rm ln}\alpha$.
Then we immediately obtain
\eq
Z_\pm(l)=-\frac{1}{8\pi^2}N_\epsilon-\frac{1}{4\pi^2}+\cdots
\en
and
\eq
{\cal M}_\Sigma={\rm Re}\biggl(2\times(\psi_C(0))^2\times T_{12}\times
(-e^2)\biggl(-\frac{1}{8\pi^2}N_\epsilon-\frac{1}{4\pi^2}\biggr)\biggr)
\en

Putting things together, we finally obtain
\eq
{\cal M}_\gamma+{\cal M}_\Sigma=
2{\rm Re}\,T_{12}\,\,\frac{\phi_0^2}{m_\pi}\,\,
\biggl(\frac{3\alpha}{4\pi}\,N_\epsilon+C_0\alpha-\frac{\alpha}{2\pi}\biggr)
\en
which induces the corresponding first-order correction in the
$\pi^+\pi^-$ atom decay width
\eq\label{roigswift}
\Gamma\rightarrow\Gamma\,\,
\biggl(\frac{3\alpha}{2\pi}\,N_\epsilon+2C_0\alpha-\frac{\alpha}{\pi}\biggr)
\en

The above expression is of course ultraviolet-divergent. It is well known
that, along with the diagrams contributing to this expression, one should
consider the four-pion Lagrangians containing (divergent) low-energy
constants in order to cure this ultraviolet divergence. This will be done
below. We would like to mention here that the term $2C_0\alpha$ from this
expression cancels with a similar term coming from the atom w.f. in
Eq. (\ref{fn}), and the final result for the decay width does not depend
on the initial approximation chosen for the Coulomb w.f. of an atom
(as it should be). Further, the term $-\alpha/\pi$ exactly
coincides with the result given in Ref. \cite{Sazdjian} obtained from
the same set of diagrams in an arbitrary covariant gauge. This provides
an independent check of the gauge invariance of our result also
for noncovariant gauges (and, in particular, for the Coulomb gauge).

Below we would like to discuss briefly the connection of our result
with the "retardation correction" given in Ref. \cite{Silagadze}.
As we have mentioned above, the matrix element ${\cal M}_\gamma$
gives exactly what is called "retardation correction". Note, however,
that our result differs somewhat from that of Ref. \cite{Silagadze}.
Namely, in this paper the virtual photon exchange diagram corresponds
to the Wick-Cutcosky model, whereas we have used the pion-photon
vertex which emerges in scalar electrodynamics. The ultraviolet divergence
which occurs in our result is a consequence of the choice of the pion-photon
vertex. Thus, strictly speaking, the present results, and the results
of Ref. \cite{Silagadze} refer to different physical models, and
cannot be directly compared. Further, as we have seen, the contribution
from ${\cal M}_\gamma$ alone is gauge-dependent and should be combined
with the self-energy diagrams to yield a gauge-invariant result.
Moreover, in gauges other than the Coulomb gauge, individual contributions
from ${\cal M}_\gamma$ and ${\cal M}_\Sigma$ contain a nonanalytic
${\rm ln}\alpha$ dependence which cancels in the sum.
Reference \cite{Silagadze} which mimicks the Feynman gauge calculations, does not
contain a nonanalytic term.

Let us now evaluate the contributions from
local four-pion Lagrangians. As we have mentioned before, it is
appropriate to include these terms, which are smooth functions of
external momenta, into the definition of $T_{12}$.
According to this convention, the transition amplitude
${\cal T}_{\pi^+\pi^-\rightarrow\pi^0\pi^0}$ from Eq. (\ref{calT})
can be written as
${\cal T}_{\pi^+\pi^-\rightarrow\pi^0\pi^0}={\cal T}_1+{\cal T}_2$,
where ${\cal T}_1$
denotes the isotopically symmetric "strong" $\pi\pi$ scattering
amplitude with the mass of the isotriplet taken equal to the {\it charged}
pion mass, and ${\cal T}_2$ includes the effect of isospin breaking as well as
the terms with low-energy constants from the four-pion Lagrangians.
From Eqs. (\ref{rel-deser}) and (\ref{roigswift}) we come to the
expression
\eq\label{elmag}
\frac{\Gamma^{(2)}}{\Gamma^{(1)}}\,\,=\,\,
\biggl(\frac{3\alpha}{2\pi}\,N_\epsilon+2C_0\alpha-\frac{\alpha}{\pi}
+2\,\frac{{\cal T}_2}{{\cal T}_1}\,\,\biggr)
\en
which displays only the electromagnetic and mass shift corrections.

As we noted before,
in Ref. \cite{Sazdjian} the amplitude ${\cal T}_2$ was evaluated at the
{\it off-mass-shell} point for $\pi^0$ mesons. However, as we see from
Eq. (\ref{rel-deser}), the amplitude emerging here is restricted on
mass shell for all external particles, and we use this prescription
hereafter. Moreover, the explicit expression for this amplitude
calculated within ChPT has become recently available \cite{Knecht},
and in the following we can use the expression given in Ref. \cite{Knecht}
as granted. All what we have to do is to extract from the amplitude
of Ref. \cite{Knecht} the terms which we have already taken into
account through the bound-state equation (vertex and self-energy corrections,
i.e. only the ones which are taken into account in the model of Roig and
Swift \cite{Roig}).

The calculations in Ref. \cite{Knecht} were carried out in the Feynman
gauge. However, as we mentioned before, the combination
$\Lambda_\gamma-\frac{e^2}{2}Z_+-\frac{e^2}{2}Z_-$ we are concerned with
is gauge-invariant, and we can safely use the results of Ref. \cite{Knecht}.
Thus, we can identify (see Eqs. (4.9)-(4.11) and (4.19) from Ref. \cite{Knecht})
\eq\label{IDENTIFY}
-2e^2(s-2m_\pi^2)G_{+-\gamma}(s)-e^2\bar J_{+-}(s)=
\Lambda_\gamma-\frac{e^2}{2}Z_+-\frac{e^2}{2}Z_-
-\frac{3e^2}{16\pi^2}N_\epsilon
-\frac{e^2}{8\pi^2}{\rm ln}\biggl(\frac{m_\pi^2}{\lambda^2}\biggr)
-\frac{e^2}{4\pi^2}
\nonumber
\en
where \cite{Knecht}
\eq
G_{+-\gamma}(s)=-i\int\frac{d^4q}{(2\pi)^4}
\frac{1}{(q^2-\lambda^2)(q^2-2q\cdot p_+)(q^2+2q\cdot p_-)},
\quad\quad p_\pm=\frac{P}{2}\pm p\nonumber\\[2mm]
J_{\alpha\beta}(l^2)=-i\int\frac{d^nq}{(2\pi)^n}
\frac{1}{(q^2-m_\alpha^2)((q-l)^2-m_\beta^2)},
\quad\quad \bar J_{\alpha\beta}(s)=J_{\alpha\beta}(s)-J_{\alpha\beta}(0)
\en
and we have introduced the photon "mass" $\lambda$ to regularize
the infrared-divergent integrals.

The amplitude ${\cal T}_{\pi^+\pi^-\rightarrow\pi^0\pi^0}$
can be easily found, subtracting
$\Lambda_\gamma-\frac{e^2}{2}Z_+-\frac{e^2}{2}Z_-$ defined by
Eq. (\ref{IDENTIFY}) from the total amplitude given in Ref.
\cite{Knecht}.
\eq
-32\pi {\cal T}_{\pi^+\pi^-\rightarrow\pi^0\pi^0}
=-\frac{s-\mo}{F^2}-B_R(s,t,u)-C_R(s,t,u)
\en
with \cite{Knecht}
\eq
B_R(s,t,u)&=&\frac{s-\mo}{F^4}\biggl[\frac{\mo}{2}\bar J_{00}(s)+
\biggl(\frac{s}{2}+2\dpi\biggr)\bar J_{+-}(s)\biggr]
\nonumber\\[2mm]
&+&\frac{1}{12F^4}\biggl[
3\biggl(t-2\mc+\frac{\dpi^2}{t}\biggr)^2
+\frac{\lambda(t,\mc,\mo)}{t^2}\biggl(\dpi^2+t(s-u)\biggr)\biggr]
\bar J_{+0}(t)
\nonumber\\[2mm]
&+&\frac{1}{12F^4}\biggl[
3\biggl(u-2\mc+\frac{\dpi^2}{t}\biggr)^2
+\frac{\lambda(u,\mc,\mo)}{u^2}\biggl(\dpi^2+u(s-t)\biggr)\biggr]
\bar J_{+0}(u)
\en
\eq\label{cr}
&&C_R(s,t,u)=
\nonumber\\[2mm]
&=&\frac{s-\mo}{32\pi^2F^4}\biggl[-\frac{\Sigma_\pi}{3}-4\dpi
-\frac{L_\pi}{\dpi}(4m_\pi^4-7\mo\mc+5m_{\pi^0}^4)
+e^2F^2(-6-6N_\epsilon+{\cal K}_1^{\pm 0})\biggr]
\nonumber\\[2mm]
&-&\frac{\mo}{32\pi^2F^4}\biggl[\frac{\mc}{3}-\frac{10\mo}{9}
-\frac{L_\pi}{\dpi}(2\mc-\mo)+\mo\bar l_3+e^2F^2{\cal K}_2^{\pm 0}\biggr]
\nonumber\\[2mm]
&+&\frac{m_\pi^4}{24\pi^2F^4}\biggl[\frac{1}{3}+L_\pi\biggr]
-\frac{\dpi}{96\pi^4F^4}\biggl[\frac{1}{t}+\frac{1}{u}\biggr]
(\Sigma_\pi\dpi-2\mo\mc L_\pi)
\nonumber\\[2mm]
&-&\frac{1}{48\pi^2F^4}\biggl[\frac{1}{6}(11s^2-t^2-u^2)
+\frac{L_\pi}{\dpi}\biggl(\biggl(\mc-\frac{3}{2}\mo\biggr)s^2
+\mo(t^2+u^2)\biggr)\biggr]
\nonumber\\[2mm]
&+&\frac{1}{48\pi^2F^4}\bar l_1(s-2\mo)(s-2\mc)
+\frac{1}{48\pi^2F^4}\bar l_2\bigl[(t-\Sigma_\pi)^2+(u-\Sigma_\pi)^2\bigr]
\en
Here
\eq
{\cal K}_1^{\pm 0}&=&\biggl(3+\frac{4Z}{9}\biggr)\bar k_1
+\frac{32Z}{9}\bar k_2+3\bar k_3+4Z\bar k_4-6L_\pi
\nonumber\\[2mm]
{\cal K}_2^{\pm 0}&=&8Z\bar k_2+3\bar k_3+4Z\bar k_4-2(1+8Z)\bar k_6-
(1-8Z)\bar k_8
\nonumber\\[2mm]
\Sigma_\pi&=&\mc+\mo,\quad\quad
\dpi=\mc-\mo,\quad\quad
L_\pi={\rm ln}\biggl(\frac{\mc}{\mo}\biggr),\quad\quad
Z=\frac{\dpi}{2e^2F^2}
\en
and $\bar l_i$, $\bar k_i$ denote the finite, renormalization scale
independent low-energy constants \cite{Knecht}. Note that in the expression
for $C_R$ the term containing the photon mass has been explicitly
cancelled (cf. Ref. \cite{Knecht}). The term proportional to
$N_\epsilon$ in this expression exactly cancels the ultraviolet divergence
which appears in Eq. (\ref{roigswift}). To demonstrate this, we note
that in the presence of this term only the lowest-order scattering amplitude
is modified as (see Eq. (\ref{cr})
\eq
-\frac{s-\mo}{F^2}\rightarrow
-\frac{s-\mo}{F^2}\biggl(1-\frac{3\alpha}{4\pi}N_\epsilon\biggr)
\en
The modification in the $\pi^+\pi^-$ atom decay width is twice as large,
and this cancels the term $\frac{3\alpha}{2\pi}N_\epsilon$ in Eq.
(\ref{roigswift}). Consequently, we can merely discard the
ultraviolet-divergent quantities simultaneously in
Eqs. (\ref{roigswift}) and (\ref{cr}).

It is convenient to expand the rest of the $\pi\pi$ scattering amplitude,
which is ultraviolet and infrared stable, in powers of $\dpi$ near
threshold. Below we present the result of this expansion retaining
only the terms proportional to $\dpi$ and $e^2F^2$ which are
sufficient up to an accuracy required. Denoting the corresponding
amplitude by ${\cal T}_{\pi^+\pi^-\rightarrow\pi^0\pi^0}^{(0)}$, we find
\eq\label{th}
&-&32\pi {\cal T}_{\pi^+\pi^-\rightarrow\pi^0\pi^0}^{(0)}
=\biggl[-\frac{3\mc}{F_\pi^2}-\frac{m_\pi^4}{32\pi^2F_\pi^4}
\biggl(11+\frac{8}{3}\bar l_1+\frac{16}{3}\bar l_2-\bar l_3+12\bar l_4\biggr)\biggr]
\nonumber\\[2mm]
&+&\dpi
\biggl[-\frac{1}{F_\pi^2}-\frac{\mc}{48\pi^2F_\pi^4}(1+4\bar l_1+3\bar l_3-12\bar l_4)\biggr]
-\frac{e^2\mc}{32\pi^2F_\pi^2}(-18+3{\cal K}_1^{\pm 0}-{\cal K}_2^{\pm 0})
+\cdots
\en
where the charged pion decay constant $F_\pi$ is related to the parameter
$F$ entering into the Lagrangian through \cite{Knecht}
\eq
F=F_\pi\left(1-\frac{\mo}{16\pi^2F_\pi^2}\bar l_4\right)
\en

The low-energy constants $\bar l_i$ and $\bar k_i$
in this equation are fixed on the
renormalization scale $\mu^2=\mo$ according to \cite{Knecht}
\eq
l_i^r(\mu)=\frac{\eta_i}{32\pi^2}
\biggl[\bar l_i+{\rm ln}\biggl(\frac{\mo}{\mu^2}\biggr)\biggr],
\quad\quad
k_i^r(\mu)=\frac{\sigma_i}{32\pi^2}
\biggl[\bar k_i+{\rm ln}\biggl(\frac{\mo}{\mu^2}\biggr)\biggr]
\en
where $\eta_i$ and $\sigma_i$ are constants. However, to
make the comparison with the calculations carried out in the isotopically
symmetric case, it is necessary to bring the normalization scale
to $\mc$. This induces the change in the second term of Eq. (\ref{th})
\eq
\frac{\mc}{48\pi^2F_\pi^4}(1+4\bar l_1+3\bar l_3-12\bar l_4)\rightarrow
\frac{\mc}{48\pi^2F_\pi^4}(1+4\bar l_1+3\bar l_3-12\bar l_4)+
\frac{19\mc}{32\pi^2F_\pi^4}
\en

After this rescaling the first term gives the isotopically symmetric
"strong" amplitude ${\cal T}_1$ and the remaining part corresponds to ${\cal T}_2$. Using
then Eqs. (\ref{elmag}) and (\ref{th}), it is easy to "read off"
the first-order mass shift and radiative corrections in
the $\pi^+\pi^-$ atom decay width
\eq\label{dm}
\delta_M&=&\frac{2\dpi}{3\mc}\biggl[1+
\frac{\mc}{48\pi^2F_\pi^2}(1+4\bar l_1+3\bar l_3-12\bar l_4)
+\frac{19\mc}{32\pi^2F_\pi^2}
\nonumber\\[2mm]
&-&\frac{\mc}{96\pi^2F_\pi^2}
\biggl(11+\frac{8}{3}\bar l_1+\frac{16}{3}\bar l_2-\bar l_3+12\bar l_4
\biggr)\biggr]
\en
\eq\label{dem}
\bar\delta_{em}&=&2C_0\alpha-\frac{\alpha}{\pi}+\frac{\alpha}{12\pi}
(-18+3{\cal K}_1^{\pm 0}-{\cal K}_2^{\pm 0})
\en

It is interesting to note that the Deser-type formula with account
of the first-order mass shift and radiative corrections can be rewritten
in a simple and transparent way. Namely, it is well known that the scattering
amplitude of charged particles develops a pole at threshold, which
corresponds to the long-range Coulomb interactions in the initial state.
Thus, at the threshold we can write \cite{Knecht}
\eq
{\rm Re} A^{+-,00}(s,t,u)=-\frac{4\mc-\mo}{F_\pi^2}\,\,\frac{e^2}{16}\,\,
\frac{m_\pi}{q}\,\,+{\rm Re}A^{+-,00}_{thr}\,\,\cdots
\en
where $q$ is the c.m.s. relative three-momentum of charged pions.

Then, the following simple
expression, valid in the lowest-order approximation, is obtained for
the $\pi^+\pi^-$ atom decay width
\eq\label{threshold}
\Gamma&=&
\frac{1}{64\pi\mc}\,\biggl(\frac{2\Delta m_\pi}{m_\pi}\biggr)^{1/2}
\biggl( 1-\frac{\Delta m_\pi}{2m_\pi}\biggr)^{1/2}
\,\,({\rm Re}A^{+-,00}_{thr})^2\,\,\phi_0^2\nonumber\\[2mm]
&\times&\biggl(1+
\biggl(-\frac{9}{8}\,\frac{\Delta E^{(1)}}{E_1}\biggr)\,
+(-2C_0\alpha)\, + \,(+2C_0\alpha)\,+\,
\left({1}/{2}+2.694-{\rm ln}\alpha\right)\frac{\Delta E^{(1)}}{E_1}\,+
\delta\,\,\biggr)
\en
where the electromagnetic and mass shift corrections, being {\it excluded}
from $\delta$, are completely taken into account in ${\rm Re}A^{+-,00}_{thr}$.
Thus, the correction factor $\delta$ displayed in Eq. (\ref{threshold}),
includes the contributions from other sources, e.g., vacuum polarization,
finite size corrections, etc. Further,
equation (\ref{threshold}) demonstrates explicitly the cancellation
of $2C_0\alpha$ terms which depend on a particular choice of the
initial approximation for the relativistic Coulomb w.f.

It is worth noting that the quantity ${\rm Re}A^{+-,00}_{thr}$ is {\it not}
proportional to the conventionally defined $\pi\pi$ scattering lengths
which acquire an additional finite contribution due to the emission of
real soft photons (see Eq. (5.17) from Ref. \cite{Knecht}). Thus Eq.
(\ref{threshold}) demonstrates, that in the presence of long-range
Coulomb force the pole-removed real part of the scattering amplitude
rather than the scattering length enters into the expression of the
first-order corrected Deser-type formula for the decay width.

Below we shall briefly discuss the comparison of our results with those
obtained in Refs. \cite{Sazdjian,Kuraev}. As we have noted before, the
main difference between our work and paper \cite{Sazdjian}, where
the corrections to the pionium decay width are also evaluated in ChPT,
consists in the fact that we argue the necessity of a different
{\it kinematic} prescription in the calculation of the $\pi\pi$ scattering
amplitude entering into the Deser-type formula. It is a completely on-mass-shell
amplitude which naturally emerges in our calculations with neutral
pions having small, but nonzero relative momentum $\vec q_0$ in the final
state. By contrast, in Ref. \cite{Sazdjian} both charged and
neutral pions have zero relative momenta, and, consequently,
neutral pions in the final state are slightly off-shell. Different
kinematic prescriptions lead to different predictions for the pionium
decay rate in Ref. \cite{Sazdjian} and the in present work.
Further, in Ref. \cite{Kuraev} the radiative corrections to the $\pi\pi$
scattering amplitude were evaluated in the Roig and Swift model, with
an explicit cutoff. The cutoff parameter was chosen to be equal to the
$\rho$-meson mass. The author also presents the calculations carried
out in the $\rho$-meson dominance model where the integrals are ultraviolet
convergent and the explicit cutoff is not needed. Thus, a direct comparison
of the results of Ref. \cite{Kuraev} with the calculations carried out
on the basis of ChPT is not possible. Note, however, that in Ref. \cite{Kuraev}
the contribution from real photon radiation is also included into the
scattering amplitude.

The mass shift and electromagnetic corrections have been evaluated in
the framework of the nonrelativistic scattering theory approach
\cite{Rasche1,Rasche2}. Strong $\pi\pi$ interactions in this approach
are described by energy-independent local potentials. It turns out that
the effect of the mass splitting in the pionium lifetime is opposite
in sign as compared to the case of field-theoretical calculations.
It is obvious that this sign depends on the choice of the "reference mass"
corresponding to the case of the isotopically symmetric world.
In particular, if the reference mass is chosen equal to the charged pion
mass \cite{Rasche2}, then this effect turns out to be negative
($\sim -3\%$). If one chooses the neutral pion mass to be the reference
mass \cite{Rasche1}, then this effect changes its sign ($\sim +3\%$).
We observe the opposite situation in our calculations based on the chiral
Lagrangian. Thus one can conclude that the energy- and mass- independent
local potentials used in Refs \cite{Rasche1,Rasche2} might not provide
an adequate description of the $\pi^+\pi^-$ system in the nonrelativistic
limit in dealing with the sophisticated issue of the isospin breaking
effect in $\pi\pi$ interactions.

\subsection{Correction due to vacuum polarization}

The vacuum polarization due to the virtual electron-positron pair
contributes in order $\alpha^2$ to the pionium decay width. However,
this effect is amplified since a small electron mass $m_e$ is present
in the denominator.

In the instantaneous approximation the photon propagator is modified
by the vacuum polarization effect as follows \cite{Efimov}
\eq
\frac{1}{\vec k^{~2}}\rightarrow\frac{1}{\vec k^{~2}}+
\frac{\alpha}{3\pi}\,\,I_{vac}(-\vec k^{~2})
\en
where
\eq\label{vacuum}
I_{vac}(-\vec k^{~2})=\int_{4m_e^2}^{\infty}\,\,
\frac{\rho(s)\, ds}{s+\vec k^{~2}},
\quad\quad
\rho(s)=\frac{1}{s}\biggl(1+\frac{2m_e^2}{s}\biggr)
\biggl(1-\frac{4m_e^2}{s}\biggr)^{1/2}
\en
The perturbation potential which is responsible for the vacuum polarization
effect is given by
\eq
V_{vac}=\frac{16i}{3}\mc\alpha^2\,\,I_{vac}(-(\vec p-\vec q\,)^2)
\en
and the corresponding matrix element from Eq. (\ref{fn}) is equal to
\eq
{\cal M}_{vac}={\rm Re}<\psi_C|V_{vac}+T_{12}G_0V_{vac}+V_{vac}G_0T_{12}
+T_{12}G_0V_{vac}G_0T_{12}|\psi_C>
\en
The first term in this expression vanishes at the bound-state energy
which is below the elastic threshold. As in the calculation of
electromagnetic radiative corrections, we neglect the fourth term
(Fig. 4d).
Thus, the matrix element can be written as follows
\eq\label{mvacuum}
{\cal M}_{vac}={\rm Re}\biggl[\frac{32}{3}\mc\alpha^2\,\, iT_{12}\,\,
\psi_C(0)\,\,\int\frac{d^4p}{(2\pi)^4}\frac{d^4q}{(2\pi)^4}G_0(\Ms;p)
I_{vac}(-(\vec p-\vec q\,)^2)\psi_C(\Ms;q)\biggr]
\en
With the use of Eq. (\ref{vacuum}) and after integrating over relative
energy variables the integral in Eq. (\ref{mvacuum}) takes the form
\eq\label{bvac}
\int_{4m_e^2}^{\infty}\, ds\,\rho(s)\,
\int\frac{d^3\vec p}{(2\pi)^3}\frac{d^3\vec q}{(2\pi)^3}\,
\frac{-i\pi\alpha m_\pi\phi_0}
{w(\vec p\,)(w(\vec q\,))^{1/2}(\vec p^{~2}+\gamma^2)
(\vec q^{~2}+\gamma^2)^2(s+(\vec p-\vec q\,)^2)}
\en

In the calculation of the integral over $d^3\vec q\,$ we can replace the
smooth factor $(w(\vec q\,))^{1/2}$ by its value at $\vec q\,=0$. Then
\eq\label{approx-3d}
\int\frac{d^3\vec q}{(2\pi)^3}\frac{1}{(w(\vec q\,))^{1/2}}
\frac{1}{(\vec q^{~2}+\gamma^2)^2\,(s+(\vec p-\vec q\,)^2)}=
\frac{1}{8\pi\gamma m_\pi^{1/2}}\,
\frac{1}{\vec p^{~2}+(\gamma+\sqrt{s})^2}+\cdots
\en

The remaining integral over $d^3\vec p\,$ can be computed analytically.
Rescaling the integration variable $s$ in Eq. (\ref{bvac}), we finally
arrive at
\eq\label{mvac}
{\cal M}_{vac}=\frac{3}{16}\alpha^2\,\,\frac{\phi_0^2}{m_e}\,\,
{\rm Re}\,T_{12}\,\,b_0
\en
where
\eq
b_0=\int_1^\infty\frac{ds(s-1)^{1/2}}{s^2}\biggl(1+\frac{1}{2s}\biggr)
f_{vac}(s)\times\Biggl[
\int_1^\infty\frac{ds(s-1)^{1/2}}{s^2}\biggl(1+\frac{1}{2s}\biggr)
\Biggr]^{-1}=0.6865\cdots
\en
\eq
f_{vac}(s)&=&\frac{2\bar m_\pi}{\pi(2\bar\gamma+\sqrt{s})}
\biggl[\,\theta(\bar m_\pi-\bar\gamma-\sqrt{s})\,\,
\frac{\bar\gamma+\sqrt{s}}{(\bar m_\pi^2-(\bar\gamma+\sqrt{s})^2)^{1/2}}
\,\,{\rm arctg}
\frac{(\bar m_\pi^2-(\bar\gamma+\sqrt{s})^2)^{1/2}}{\bar\gamma+\sqrt{s}}
\nonumber\\[2mm]
&+&\theta(\sqrt{s}-\bar m_\pi+\bar\gamma)\,\,
\frac{\bar\gamma+\sqrt{s}}{((\bar\gamma+\sqrt{s})^2-\bar m_\pi^2)^{1/2}}
\,\,{\rm ln}\biggl(\,\frac
{\bar\gamma+\sqrt{s}+((\bar\gamma+\sqrt{s})^2-\bar m_\pi^2)^{1/2}}
{\bar\gamma+\sqrt{s}-((\bar\gamma+\sqrt{s})^2-\bar m_\pi^2)^{1/2}}
\,\biggr)\nonumber\\[2mm]
&-&\frac{\bar\gamma}{(\bar m_\pi^2-\bar\gamma^2)^{1/2}}\,\,
{\rm arctg}\,
\frac{(\bar m_\pi^2-\bar\gamma^2)^{1/2}}{\bar\gamma}
\biggr]
\en
and $\bar m_\pi=m_\pi/(2m_e)$, $\bar\gamma=\gamma/(2m_e)$.

Using the relativistic Deser formula in the lowest-order approximation
(\ref{rel-deser}), it is easy to observe that Eq. (\ref{mvac}) leads
to the following modification in the $\pi^+\pi^-$ atom decay width
due to the vacuum polarization effect
\eq\label{vac-mod}
\Gamma\rightarrow\Gamma\biggl(1+\frac{3}{16}\,\alpha^2\,\,
\frac{m_\pi}{m_e}\,\, b_0\biggr)
\en

Note that in Ref. \cite{Efimov} the vacuum polarization correction to the
pionium lifetime was calculated only with account of discrete
spectrum transitions. Thus, our result is a generalization
of that from Ref. \cite{Efimov}.

The NRQED based calculation of the vacuum polarization effect
in the pionium lifetime has become available recently \cite{Labelle}. We find
that the analytic expression of the so-called 0-Coulomb term in
Ref. \cite{Labelle} coincides with our result up to the relativistic
kinematic factor in the w.f. whose presence is due to the choice
of Barbieri-Remiddi kernel. Numerically the effect of this factor, which
contributes in higher orders in $\alpha$, is very small.
In Ref. \cite{Labelle} the results for 1-Coulomb and multi-Coulomb
contributions are also given. These contributions, which formally are of
higher order in $\alpha$, would emerge in our calculations as
second-order perturbative corrections to the pionium lifetime.

\subsection{Finite size correction}

In the presence of a pion loop the bare VPP vertex is modified. According
to Ref. \cite{ChPT},
\eq\label{vertex}
F_V(t)=1+\frac{1}{6F_\pi^2}(t-4\mc)\bar J_{+-}(t)
+\frac{t}{96\pi^2F_\pi^2}\biggl(\bar l_6-\frac{1}{3}\biggr)
=1+\delta F_V(t)
\en
In the instantaneous approximation the perturbation potential is given
by (see Fig. 5)
\eq
V_F=8ie^2\mc\,\,\frac{\delta F_V[-(\vec p-\vec q\,)^2]}{(\vec p-\vec q\,)^2}
\en
and the corresponding matrix element reads as
\eq
{\cal M}_F=<\psi_C|V_F+T_{12}G_0V_F+V_FG_0T_{12}+T_{12}G_0V_FG_0T_{12}|\psi_C>
\en
Again, the first term in this matrix element vanishes at the bound-state
energy, and we neglect the fourth term. The remainder is then given by
\eq\label{calmf}
{\cal M}_F={\rm Re}\,\biggl[2\,\psi_C(0)\, T_{12}\,\,
\int\frac{d^4p}{(2\pi)^4}\frac{d^4q}{(2\pi)^4}G_0(\Ms;p)
\frac{8ie^2m_\pi^2\delta F_V[-(\vec p-\vec q\,)^2]}{(\vec p-\vec q\,)^2}
\psi_C(\Ms;q)\biggr]
\en
The integral over $d^4p$ diverges in the ultraviolet region. This stems
from the fact that the diagram depicted in Fig. 6a, is ultraviolet-divergent
as a whole, though the subdivergence in the VPP vertex has already been
removed by an appropriate counterterm depicted in Fig. 6b (this counterterm
is implicit in Eqs. (\ref{vertex}) and (\ref{calmf})). Thus, a higher-order
counterterm shown in Fig. 6c is needed to cancel the overall divergence
in Fig. 6a (and, consequently, in Eq. (\ref{calmf})). It is obvious that
this divergence is removed analogously to that from subsection
D (Mass shift and radiative corrections), and we shall not further dwell
upon this question.

To simplify the calculations in the
relevant low-$t$ region, instead of Eq. (\ref{vertex}) we use the well-known
monopole parameterization
\eq\label{monopole}
\delta F_V(t)=\frac{t}{\mu_V^2-t},\quad\quad
\mu_V^{-2}=\frac{1}{6}<r^2>_V^\pi
\en
with the same $<r^2>_V^\pi$ as in Eq. (\ref{vertex}). The integral
in Eq. (\ref{calmf}) is then convergent and can be easily evaluated,
resulting in
\eq\label{finite-3d}
-2e^2\mc\,\,\int\frac{d^3\vec p}{(2\pi)^3}\frac{d^3\vec q}{(2\pi)^3}\,\,
\frac{4\pi\alpha m_\pi\phi_0}
{w(\vec p\,)(w(\vec q\,))^{1/2}(\vec p^{~2}+\gamma^2)(\vec q^{~2}+\gamma^2)^2
(\mu_V^2+(\vec p-\vec q\,)^2)}
\en
With the use of Eq. (\ref{approx-3d}) and the inequalities
$\mu_V>>m_\pi>>\gamma$ the integration of Eq. (\ref{finite-3d}) gives
\eq\label{finite-int}
-2e^2\mc\,\frac{\phi_0}{m_\pi^{1/2}}\,
\int\frac{d^3\vec p}{(2\pi)^3}\,
\frac{1}{w(\vec p\,)(\vec p^{~2}+\gamma^2)(\vec p^{~2}+\mu_V^2)}
=\frac{\phi_0}{m_\pi^{1/2}}\,
\frac{e^2\mc}{2\pi^2\mu_V^2}\,
{\rm ln}\,\frac{4\mu_V^2}{\mc}+\cdots
\en
and
\eq
{\cal M}_F=-\frac{\phi_0^2}{m_\pi}\,{\rm Re}T_{12}\,\,
\frac{4\alpha}{\pi}\,\frac{\mc}{\mu_V^2}\,\,{\rm ln}\,
\frac{4\mu_V^2}{\mc}+\cdots
\en
The modification of the $\pi^+\pi^-$ atom decay width due to the finite
size effect is given by
\eq\label{finitesize}
\Gamma\rightarrow\Gamma\biggl(1-
\frac{4\alpha}{\pi}\,\frac{\mc}{\mu_V^2}\,\,{\rm ln}\,
\frac{4\mu_V^2}{\mc}\biggr)
\en

\section{Numerical results and discussion}

In this section we present the numerical results on the lowest-order
corrections to the pionium decay width. To this end we combine various
corrections obtained in the previous section. As we have seen, the
correction due to the relativistic modification of Coulomb w.f.
cancels with the corresponding piece in electromagnetic radiative
corrections. In the final result we cancel these corrections explicitly.

Below we give a list of analytic results on the lowest-order
corrections to the $\pi^+\pi^-$ atom decay width
\eq\label{final}
\Gamma&=&\frac{16\pi}{9}\,
\biggl(\frac{2\,\Delta m_\pi}{m_\pi}\biggr)^{{1}/{2}}
\biggl(1-\frac{\Delta m_\pi}{2m_\pi}\biggr)^{1/2}
(a_0^0-a_0^2)^2\,\,\phi_0^2\,\,
(1+\delta_S+\delta_C+\delta_M+\delta_{em}+\delta_{vac}+\delta_F)
\nonumber\\&&
\en
where $a_0^0$ and $a_0^2$ denote the $\pi\pi$ scattering lengths
in the isospin-symmetric case, with the charged pion mass taken
to be the common mass of the pion isotriplet.

\noindent
$\delta_S$ is the correction due to the displacement of the
bound-state pole by strong interactions (see Eq. (\ref{fn}))
\eq
\delta_S=-\frac{9}{8}\,\,\frac{\Delta E^{(1)}}{E_1}=
-5.47\times 10^{-3}\,\,m_\pi(2a_0^0+a_0^2)
\en
$\delta_C$ corresponds to the correction due to the Coulomb photon exchanges
(Eq. (\ref{fn}))
\eq
\delta_C=(1/2+2.694-{\rm ln}\alpha)\frac{\Delta E^{(1)}}{E_1}=
3.95\times 10^{-2}\,\,m_\pi(2a_0^0+a_0^2)
\en
$\delta_M$ stands for the correction due to the $m_{\pi^\pm}-m_{\pi^0}$
mass difference (Eq. (\ref{dm})) and $\delta_{em}$ corresponds to the
electromagnetic corrections without $2C_0\alpha$ term (Eq. (\ref{dem}).
The quantity $\delta_{vac}$ denotes the correction due to the vacuum polarization
effect (Eq. (\ref{vac-mod}))
\eq
\delta_{vac}=\frac{3}{16}\,\alpha^2\,\frac{m_\pi}{m_e}\times 0.6865
\en
and $\delta_F$ corresponds to the finite size correction
(Eq. (\ref{finitesize}))
\eq
\delta_F=\frac{2\alpha}{3\pi}\,\,\mc<r^2>_V^\pi\,\,
{\rm ln}\,\biggl(\frac{1}{24}\,\mc<r^2>_V^\pi\biggr)
\en

To make numerical estimation of the above-listed corrections,
one has to substitute the values of low-energy constants into
these expressions. For the constants $\bar l_i$ we take the
numerical values from Ref. \cite{ChPT}
$\bar l_1=-2.3\pm 2.7$, $\bar l_2=6.0\pm 1.3$, $\bar l_3=2.9\pm 2.4$,
$\bar l_4=4.3\pm 0.9$.
The constants $\bar k_i$ are more difficult to estimate. In our paper
we use the values from Ref. \cite{Knecht} based on a rough
estimate at the scale coinciding with the $\rho$-meson mass
$|k_i^r(m_\rho)|\leq\frac{1}{16\pi^2}$. This estimate yields
$\frac{e^2F_\pi^2}{\mc}\,{\cal K}_1^{\pm 0}=1.8\pm 0.9$,
$\frac{e^2F_\pi^2}{\mc}\,{\cal K}_2^{\pm 0}=0.5\pm 2.2$ \cite{Knecht}.
Large error bars in the low-energy constants
${\cal K}_1^{\pm 0},~{\cal K}_2^{\pm 0}$, in turn, do not allow
one to calculate the electromagnetic radiative correction to the
atom decay width with a high accuracy.
Other input parameters in our calculations are the S-wave $\pi\pi$
scattering lengths: $a_0^0=0.217m_\pi^{-1}$, $a_0^2=-0.041m_\pi^{-1}$
calculated in ChPT and the e.m. charge radius of pion
$<r^2>_V^\pi=0.439~Fm$ \cite{ChPT}.

Substituting the above values of the input parameters into the expressions
for various corrections to the decay width we obtain
the results collected in Table I. As we observe from this table,
the largest correction in the decay width is caused by the mass
splitting effect in accordance with the result of
Refs. \cite{Rasche1,Rasche2,Sazdjian}. Our result for this effect has the
same sign as the result of Ref. \cite{Sazdjian}, but is different in
magnitude. The reason for this difference is traced back to
different kinematic prescriptions used for the $\pi\pi$ scattering
amplitude in the Deser formula (see the discussion in the text).
The sign of the mass splitting effect obtained in the nonrelativistic
scattering theory approach \cite{Rasche1,Rasche2}, turns out to be opposite
as compared to our result, and is of the same order of magnitude.

Our last remark concerns the effect of the $m_d-m_u$ mass splitting
on the pionium decay width. It is well known that in the one loop
order this leads only to the shift in the neutral pion mass \cite{ChPT}.
Since in our calculations of the on-shell $\pi\pi$ scattering amplitude
we use the physical values of the pion masses,
with the mass difference caused both by $m_d-m_u\neq 0$ and electromagnetic
corrections, the resulting mass splitting correction includes both these
effects.
\footnote{
We are indebted to Prof. J. Gasser for the clarifying discussions on this
problem.
}

\vspace*{.4cm}

{\it Acknowledgments}.
We thank V. Antonelli, A. Gall, A. Gashi, J. Gasser,
E.A. Kuraev, H. Leutwyler, P. Minkowski, L.L. Nemenov, E. Pallante
and Z. Silagadze
for useful
discussions, comments and remarks. A.G.R. thanks Bern University for
the hospitality where part of this work was completed.
This work was supported in part by the Russian Foundation for
Basic Research (RFBR) under contract 96-02-17435-a.

\appendix

\section{}
\setcounter{equation}{0}
\def\theequation{A\arabic{equation}}

In this appendix, we present the calculation of various integrals
which appear in the first-order correction terms. We start from the
evaluation of the relativistic Coulomb w.f. at the origin
(Eq. (\ref{c0})). By carrying out the $p_0$ integration with the use of
Cauchy theorem, this quantity can be written as
\eq\label{A1}
\psi_C(0)&=&\int\frac{d^4p}{(2\pi)^4}\psi_C(\Ms;p)=
\frac{\phi_0}{m_\pi^{1/2}}\int\frac{d^3\vec p}{(2\pi)^3}
\biggl(\frac{m_\pi}{w(\vec p\,)}\biggr)^{1/2}
\frac{4\pi\alpha m_\pi}{(\vec p^{~2}+\gamma^2)^2}
\nonumber\\[2mm]
&=&
\frac{\phi_0}{m_\pi^{1/2}}\biggl(1-\int\frac{d^3\vec p}{(2\pi)^3}
\frac{\vec p^{~2}}{(w(\vec p\,))^{1/2}(m_\pi^{1/2}+(w(\vec p\,))^{1/2})
(m_\pi+w(\vec p\,))}\,
\frac{4\pi\alpha m_\pi}{(\vec p^{~2}+\gamma^2)^2}\biggr)
\en
In the lowest-order approximation in $\alpha$ we replace the factor
$(\vec p^{~2}+\gamma^2)^2$ in the denominator by $(\vec p^{~2})^2$
and obtain
\eq\label{A2}
\psi_C(0)=\frac{\phi_0}{m_\pi^{1/2}}(1-C_0\alpha)+\cdots
\en
where
\eq\label{A3}
C_0=\frac{2}{\pi}\int_0^\infty\frac{dp}
{(1+p^2)^{1/4}(1+(1+p^2)^{1/4})(1+(1+p^2)^{1/2})}=0.381\cdots
\en

Next we turn to the calculation of the integral which is present in
Eq. (\ref{AA}). This integral contains the Green  function $\delta G(p,q)$
which corresponds to the exchanged Coulomb photon ladders and, according
to the Eqs. (\ref{REGULAR}), (\ref{TILDE}), is given by
\eq\label{REG}
\delta G(p,q)&=&
i\bigr( w(\vec p\,)w(\vec q\,)\bigl)^{1/2}
\biggl[\Phi(\vec p,\vec q\,)-
S(\vec p\,)S(\vec q\,)\frac{8}{\Ms}\frac{\partial}{\partial\Ms}\biggr]
G_0(\Ms;p)G_0(\Ms;q)
\nonumber\\[2mm]
\Phi(\vec p,\vec q\,)&=&16\pi m_\pi\alpha\biggl[
\frac{1}{(\vec p-\vec q\,)^2}+I_R(\vec p,\vec q\,)\biggr]
+\frac{1}{(m_\pi\alpha)^2}S(\vec p\,)S(\vec q\,)R(\vec p,\vec q\,)
\\[2mm]
R(\vec p,\vec q\,)&=&25-\biggl(\frac{8}{\pi m_\pi\alpha}\biggr)^{1/2}
\bigl[S(\vec p\,)+S(\vec q\,)\bigr]+\cdots
\nonumber
\en
where ellipses stand for the higher-order terms in $\alpha$.

Substituting this expression in the integral from Eq. (\ref{AA}) and carrying
out $p_0$, $q_0$ integrations, we obtain
\eq\label{I1}
&&
I_1=\int\frac{d^4p}{(2\pi)^4}\frac{d^4q}{(2\pi)^4}\bar\psi_C(\Ms;p)
(G_0^{-1}(\Ms;p))'\delta G(p,q)=
\\[2mm]
&&
\int\frac{d^3\vec p}{(2\pi)^3}\frac{d^3\vec q}{(2\pi)^3}
\frac{4\pi\alpha m_\pi\phi_0}{(\vec p^{~2}+\gamma^2)^3}
\biggl[\Phi(\vec p,\vec q\,)-S(\vec p\,)S(\vec q\,)
\biggl(\frac{6}{\vec p^{~2}+\gamma^2}+\frac{4}{\vec q^{~2}+\gamma^2}
\biggr)\biggr]\frac{-\Ms}{8(w(\vec q\,))^{1/2}}
\frac{1}{\vec q^{~2}+\gamma^2}
\nonumber
\en
where we have used
\eq
\int\frac{dp_0}{2\pi i}\,(G_0(\Ms;p))'(G_0^{-1}(\Ms;p))'G_0(\Ms;p)=
\frac{3(\Ms)^2}{32w(\vec p\,)(\vec p^{~2}+\gamma^2)^3}+\cdots
\en
In the calculation of 3D integrals containing the function
$\Phi(\vec p,\vec q\,)$, we use
\eq\label{A7}
\int\frac{d^3\vec p}{(2\pi)^3}\frac{1}{(\vec p-\vec q\,)^2}
\frac{1}{(\vec p^{~2}+\gamma^2)^3}=
\frac{1}{4\pi\alpha^3 m_\pi^3}\frac{1}{(\vec q^{~2}+\gamma^2)}+
\frac{1}{8\pi\alpha m_\pi}\frac{1}{(\vec q^{~2}+\gamma^2)^2}
\en
and
\eq\label{A8}
\int\frac{d^3\vec p}{(2\pi)^3}\,\,\frac{I_R(\vec p,\vec q\,)}
{(\vec p^{~2}+\gamma^2)^3}=
\int_0^1\frac{d\rho}{\rho}
({\cal J}_1'(\rho;\vec q\,)-{\cal J}_1''(\rho;\vec q\,))
\en
where
\eq
{\cal J}_1'(\rho;\vec q\,)=
\int\frac{d^3\vec p}{(2\pi)^3}
\frac{1}{(\vec p^{~2}+\gamma^2)^3}\,\,
\frac{1}{(\vec p-\vec q\,)^2\rho+
m_\pi^{-2}\alpha^{-2}(\vec p^{~2}+\gamma^2)(\vec q^{~2}+\gamma^2)(1-\rho)^2}
\en
The integration over $d^3\vec p$ can be carried out with the use of
Feynman parameterization. We obtain
\eq\label{A10}
{\cal J}_1'(\rho;\vec q\,)&=&
\frac{3}{8\pi\alpha^3 m_\pi^3}\,
\frac{1}{(\vec q^{~2}+\gamma^2)}
\int_0^1\frac{dx(1-x)^2}{d_-^{3/2}d_+^{5/2}}+
\frac{3}{32\pi\alpha m_\pi}\,
\frac{1}{(\vec q^{~2}+\gamma^2)^2}
\int_0^1\frac{dx(1-x)^2}{d_-^{5/2}d_+^{5/2}}
\nonumber\\[2mm]
d_\pm&=&1-x+\frac{x}{4}(1\pm\rho)^2
\en
and
\eq\label{A11}
{\cal J}_1''(\rho;\vec q\,)=
\int\frac{d^3\vec p}{(2\pi)^3}
\frac{1}{(\vec p^{~2}+\gamma^2)^3}\,\,
\frac{1}{m_\pi^{-2}\alpha^{-2}(\vec p^{~2}+\gamma^2)(\vec q^{~2}+\gamma^2)}
=\frac{1}{2\pi\alpha^3 m_\pi^3}\,\frac{1}{(\vec q^{~2}+\gamma^2)}
\en
Substituting Eqs. (\ref{A11}) and (\ref{A10}) into (\ref{A8}) and carrying
out the integration over $dx$ and $d\rho$, we finally obtain
\eq\label{A12}
\int\frac{d^3\vec p}{(2\pi)^3}\,\,\frac{I_R(\vec p,\vec q\,)}
{(\vec p^{~2}+\gamma^2)^3}=
-\frac{1}{4\pi\alpha^3 m_\pi^3}\,
\frac{1}{(\vec q^{~2}+\gamma^2)}+
\frac{1}{8\pi\alpha m_\pi}\,
\frac{1}{(\vec q^{~2}+\gamma^2)^2}
\en
With the use of Eqs. (\ref{REG}), (\ref{A7}) and (\ref{A12}) the integration
in Eq. (\ref{I1}) is trivially carried out, resulting in
\eq
I_1=\frac{\phi_0}{m_\pi^{1/2}}\,\,\frac{1}{\alpha^2 m_\pi^2}+\cdots
\en
Substituting this result back in Eq. (\ref{AA}), we readily obtain the
final result given in this equation.

Next we turn to the calculation of the integral which is present in
Eq. (\ref{BB})
\eq
I_2=\int\frac{d^4p}{(2\pi)^4}\frac{d^4q}{(2\pi)^4}\delta G(p,q)=I_2'+I_2''
\en
where $I_2'$ and $I_2''$ correspond to the "nonderivative" and "derivative"
terms in Eq. (\ref{REG}). Carrying out the integration over the relative
energies with the use of Cauchy theorem, $I_2'$ can be written as
\eq
I_2'=-\frac{i}{16}\int\frac{d^3\vec p}{(2\pi)^3}\frac{d^3\vec q}{(2\pi)^3}\,\,
\frac{1}{(w(\vec p\,)w(\vec q\,))^{1/2}}\,\,
\frac{1}{(\vec p^{~2}+\gamma^2)(\vec q^{~2}+\gamma^2)}\,\,\Phi(\vec p,\vec q\,)
\en
$I_2'$ receives the contribution from (a) one-photon exchange, (b) multiphoton
exchanges concentrated in $I_R(\vec p,\vec q\,)$ and (c) the rest, proportional
to the function $R$ (see Eq. (\ref{REG}). Below we shall evaluate these
contributions separately.

\noindent (a) One Coulomb photon exchange
\eq\label{A15}
I_{2a}'=\int\frac{d^3\vec p}{(2\pi)^3}
\frac{-i\pi\alpha m_\pi}{(w(\vec p\,)w(\vec q\,))^{1/2}}\,\,
\frac{1}{(\vec p^{~2}+\gamma^2)(\vec q^{~2}+\gamma^2)}\,\,
\frac{1}{(\vec p-\vec q\,)^2}=
-\frac{i\alpha m_\pi}{4}\int d^3\vec r\,\,\frac{\varphi^2(r)}{r}
\en
where
\eq\label{A16}
\varphi(r)=\int\frac{d^3\vec p}{(2\pi)^3}\,e^{-i\vec p\vec r}
\,\,\frac{1}{(w(\vec p\,))^{1/2}(\vec p^{~2}+\gamma^2)}
\en
Using exponential parameterization, the integration over $d^3\vec p$
in Eq. (\ref{A16}) can be carried out, resulting in
\eq\label{A17}
\varphi(r)=\frac{1}{8\pi^{3/2}\Gamma(\frac{1}{4})}
\int_0^\infty duu^{-5/4}\int_0^1 dxx^{-3/4}
\exp\biggl[-\frac{\vec r^{~2}}{4u}-u(1-x)\gamma^2-ux\mc\biggr]
\en
Substituting Eq. (\ref{A17}) into Eq. (\ref{A15}) and integrating,
we obtain
\eq
I_{2a}'=-\frac{i\alpha m_\pi}{32\pi^{3/2}\Gamma^{2}(\frac{1}{4})}
\int_0^1d\tau\int_0^1dx_1\int_0^1dx_2\,\,
\frac{\tau^{-1/4}(1-\tau)^{-1/4}x_1^{-3/4}x_2^{-3/4}}
{(\gamma^2+(\mc-\gamma^2)(\tau x_1+(1-\tau)x_2))^{1/2}}
\en
Note that one can not directly assume here $\gamma=0$ in the denominator,
since the integral over the Feynman parameters diverges in this limit.
In order to overcome this difficulty, we split the integration area into
two domains according to
\eq
&&
\int_0^1dx_1\int_0^1dx_2f(x_1,x_2)=
\int_0^1\rho d\rho\int_0^1 dtf(\rho t,\rho(1-t))+
\int_1^2\rho d\rho\int_{1-1/\rho}^{1/\rho}dtf(\rho t,\rho(1-t))
\nonumber\\
\en
Performing explicitly the integration over $d\rho$ in the first
domain, we obtain
\eq
&&I_{2a}'(1)=
-\frac{i\alpha m_\pi}{16\pi^{3/2}\Gamma^2(\frac{1}{4})(\mc-\gamma^2)^{1/2}}
\int_0^1dt\int_0^1d\tau
\frac{\tau^{-1/4}(1-\tau)^{-1/4}t^{-3/4}(1-t)^{-3/4}}
{(\tau t+(1-\tau)(1-t))^{1/2}}\nonumber\\[2mm]
&&\times {\rm ln}\,\,\bigl[\gamma^{-1}\bigl(
(\mc-\gamma^2)^{1/2}(\tau t+(1-\tau)(1-t))^{1/2}+
(\gamma^2+(\mc-\gamma^2)(\tau t+(1-\tau)(1-t)))^{1/2}\bigr)\bigr]
\nonumber\\[2mm]
&&=\frac{i\alpha{\rm ln}\alpha}{16\pi}
-\frac{i\alpha}{16\pi}\biggl[ 2{\rm ln}2+\frac{c_1}{2\pi^{1/2}\Gamma^2(1/4)}
\biggr]+\cdots
\en
where ellipses stand for the higher order terms in $\alpha$ and
\eq
&&
c_1=\int_0^1dt\int_0^1d\tau\,\,
\frac{\tau^{-1/4}(1-\tau)^{-1/4}t^{-3/4}(1-t)^{-3/4}}
{(\tau t+(1-\tau)(1-t))^{1/2}}
{\rm ln}(\tau t+(1-\tau)(1-t))=-40.374\cdots\nonumber\\
&&
\en
The second integral converges when $\gamma\rightarrow 0$ in the denominator,
resulting in
\eq
I_{2a}'(2)=-\frac{i\alpha c_2}{4\pi^{3/2}\Gamma^2(\frac{1}{4})}
\en
where
\eq
&&
c_2=\frac{1}{4}\int_0^1d\tau\tau^{-1/4}(1-\tau)^{-1/4}
\int_1^2d\rho\,{\rm ln}\rho\,(\rho-1)^{-3/4}(\tau+(1-\tau)(\rho-1))^{-1/2}
=0.288\cdots\nonumber\\
&&
\en

(b) Multiphoton exchanges

In this contribution we can safely replace the smooth factor in the
denominator $(w(\vec p\,)w(\vec q\,))^{1/2}\rightarrow m_\pi$. Then
\eq
I_{2b}=-i\pi\alpha\int\frac{d^3\vec p}{(2\pi)^3}\frac{d^3\vec q}{(2\pi)^3}
\frac{I_R(\vec p,\vec q\,)}{(\vec p^{~2}+\gamma^2)(\vec q^{~2}+\gamma^2)}=
-i\pi\alpha\int_0^1\frac{d\rho}{\rho}{\cal J}_{2b}'(\rho)
\en
where according to Eq. (\ref{REGULAR})
\eq
&&{\cal J}_{2b}'(\rho)=
\int\frac{d^3\vec p}{(2\pi)^3}\frac{d^3\vec q}{(2\pi)^3}
\frac{1}{(\vec p^{~2}+\gamma^2)(\vec q^{~2}+\gamma^2)}
\nonumber\\[2mm]
&\times&\biggl[\frac{1}{(\vec p-\vec q\,)^2\rho+
m_\pi^{-2}\alpha^{-2}(\vec p^{~2}+\gamma^2)(\vec q^{~2}+\gamma^2)(1-\rho)^2}
-
\frac{1}{m_\pi^{-2}\alpha^{-2}(\vec p^{~2}+\gamma^2)(\vec q^{~2}+\gamma^2)}
\biggr]
\en
Introducing Feynman parameters and carrying out the momentum integration,
we obtain
\eq
{\cal J}_{2b}'(\rho)=\frac{1}{16\pi^2}
\biggl[\frac{1}{2}\int_0^1\frac{dx}{d_-d_+^{1/2}(d_-^{1/2}+d_+^{1/2})}-1\biggr]
\en
and
\eq
I_{2b}'=-\frac{i\alpha}{16\pi}
\en

(c) Factorizing integrals

The integral containing the function $R$ is evaluated in the straightforward
manner, resulting in
\eq
I_{2c}'=-\frac{17i\alpha}{128\pi}+\cdots
\en

The "derivative" term $I_2''$ can be easily calculated. The integration in the
variables $p$ and $q$ again factorizes, and we have
\eq
I_2''=\frac{i\alpha}{16\pi}+\cdots
\en
Putting all together, we finally arrive at the result
\eq
I_2=\frac{i\alpha}{16\pi}{\rm ln}\alpha+
\frac{i\alpha}{16\pi}\biggl[-\frac{17}{8}-2{\rm ln}2-
\frac{c_1}{2\pi^{1/2}\Gamma^2(\frac{1}{4})}-
\frac{4c_2}{\pi^{1/2}\Gamma^2(\frac{1}{4})}\biggr]=
\frac{i\alpha}{16\pi}({\rm ln}{\alpha}-2.694)
\en
which is substituted in Eq. (\ref{BB}).

\section{}
\setcounter{equation}{0}
\def\theequation{B\arabic{equation}}

In this appendix we shall present the calculation of the integrals appearing
in the electromagnetic radiative corrections (Eq. (\ref{em-radiative})).
\eq
&&
\int\frac{d^4p}{(2\pi)^4}\frac{d^4q}{(2\pi)^4}
G_0(\Ms;p)\frac{ie^2((\Ms)^2-(p_0+q_0)^2)}{(\vec p-\vec q\,)^2}
G_0(\Ms;q)\,\, 4i(w(\vec q\,))^{1/2}
\frac{4\pi\alpha m_\pi\phi_0}{\vec q^{~2}+\gamma^2}=
\tilde I_1+\tilde I_2\nonumber\\
&&
\en
Integrating over the relative energy variables, the first term is
rewritten in the form
\eq
\tilde I_1&=&\frac{e^2(\Ms)^2}{4}\int\frac{d^3\vec p}{(2\pi)^3}
\frac{1}{w(\vec p\,)(\vec p^{~2}+\gamma^2)}
\int\frac{d^3\vec q}{(2\pi)^3}\frac{1}{(w(\vec q\,))^{1/2}(\vec p-\vec q\,)^2}
\frac{4\pi\alpha m_\pi\phi_0}{(\vec q^{~2}+\gamma^2)^2}
\nonumber\\[2mm]
&=&\frac{e^2\phi_0(\Ms)^2}{4m_\pi^{1/2}}
\int\frac{d^3\vec p}{(2\pi)^3}
\frac{1}{w(\vec p\,)(\vec p^{~2}+\gamma^2)^2}+\cdots
\en
Using the same trick as in Eqs. (\ref{A1})-(\ref{A3}), we can write
\eq
\tilde I_1=\frac{\phi_0}{m_\pi^{1/2}}
\biggl(1-\frac{e^2}{2\pi^2}\int_0^\infty
\frac{dp}{(1+p^2)^{1/2}(1+(1+p^2)^{1/2})}+\cdots\biggr)=
\frac{\phi_0}{m_\pi^{1/2}}\biggl(1-\frac{2\alpha}{\pi}\biggr)+\cdots
\en

In the calculations of $\tilde I_2$ the term containing $2p_0q_0$ vanishes
since it is odd in $p_0$ and $q_0$. Thus one can write
$\tilde I_2=\tilde I_2'+\tilde I_2''$ where $\tilde I_2'$ and $\tilde I_2''$
contain $p_0^2$ and $q_0^2$, respectively.

$\tilde I_2'$ is ultraviolet divergent. Introducing dimensional regularization,
we can write
\eq
&&\int\frac{d^4p}{(2\pi)^4}G_0(\Ms;p)\frac{p_0^2}{(\vec p-\vec q\,)^2}
\rightarrow\nonumber\\[2mm]
&\rightarrow&-(\mu^2)^{2-n/2}\int\frac{d^np}{(2\pi)^n}
\frac{p_0^2}{\biggl(\bigl(\frac{P}{2}+p\bigr)^2-\mc\biggr)
             \biggl(\bigl(\frac{P}{2}-p\bigr)^2-\mc\biggr)
             (\vec p-\vec q\,)^2}
\nonumber\\[2mm]
&=&\frac{i}{16\pi^2}\biggl(N_\epsilon+4
+\frac{w(\vec q\,)}{|\vec q\,|}{\rm ln}\frac
{w(\vec q\,)-|\vec q\,|}{w(\vec q\,)+|\vec q\,|}\biggr)
\en
and
\eq
\tilde I_2'=\frac{\alpha}{4\pi}(N_\epsilon+2)\,\,\frac{\phi_0}{m_\pi^{1/2}}
\en

$\tilde I_2''$ does not contain the ultraviolet divergence. Integrating
over relative energy variables, we obtain
\eq
\tilde I_2''=\frac{e^2}{4}\int\frac{d^3\vec p}{(2\pi)^3}\frac{d^3\vec q}{(2\pi)^3}
\frac{1}{w(\vec p\,)(w(\vec q\,))^{1/2}}
\frac{1}{(\vec p-\vec q\,)^2(\vec p^{~2}+\gamma^2)}
\frac{4\pi\alpha m_\pi\phi_0}{(\vec q^{~2}+\gamma^2)}
\en
It is easy to see that $\tilde I_2''$ leads to the modification of the
$\pi^+\pi^-$ decay width in the order $\alpha^2{\rm ln}\alpha$ and
thus can be safely neglected. The final result reads as (cf. with
Eq. (\ref{em-radiative}))
\eq
\tilde I=\tilde I_1+\tilde I_2=
\frac{\phi_0}{m_\pi^{1/2}}\biggl(1+\frac{\alpha}{4\pi}N_\epsilon
-\frac{3\alpha}{2\pi}\biggr)
\en

\newpage

\begin{center}
{\bf Table I. Corrections to the $\pi^+\pi^-$ atom decay width}
\vspace{1.5cm}

\def\arraystretch{1.5}
\begin{tabular}{|l|c|c|}
\hline
~~Effect                    & ~~Value~~      & ~~Correction (in \%)~~ \\
\hline
~~Strong                    & $\delta_S$     & $-0.22$                \\
\hline
~~Coulomb photon exchange~~ & $\delta_C$     & $+1.55$                \\
\hline
~~Mass shift                & $\delta_M$     & $+2.99\pm 0.77$          \\
\hline
~~Electromagnetic radiative & $\delta_{em}$  & $+1.73\pm 2.31$         \\
\hline
~~Vacuum polarization       & $\delta_{vac}$ & $+0.19$                \\
\hline
~~Finite size               & $\delta_F$     & $-0.16$                \\
\hline
~~Total                     & $\delta_{tot}$ & $+6.1\pm 3.1$        \\
\hline
\end{tabular}
\end{center}

\newpage
\centerline{{\bf FIGURE CAPTIONS}}
\vspace*{.7cm}

\noindent {\bf Fig. 1}.
Diagrammatic representation of the Bethe-Salpeter equation for
the $\pi^+\pi^-$ atom w.f. Initial equation (a) through the redefinition
of the kernel (b) and the w.f. is transformed into the equation (c).
The new kernel $V$ contains the self-energy insertions in the
{\it outgoing} external lines only (see (b)).

\noindent {\bf Fig. 2}.
Matrix element describing the "residual" photon exchange. The imaginary
part of the diagram (a) vanishes at the bound-state energy. Diagram (d)
which is of the second order in the "strong" amplitude $T_{12}$, is neglected
in the present approximation. Dashed line denotes the virtual photon
propagator, and dots correspond to the instantaneous virtual photon
exchange.

\noindent {\bf Fig. 3}.
Matrix element corresponding to the self-energy correction in the
external pion legs (Eq. (\ref{fig3})).

\noindent {\bf Fig. 4}.
Vacuum polarization correction.

\noindent {\bf Fig. 5}.
Vertex correction.

\noindent {\bf Fig. 6}.
Cancellation of the divergences which are present in the expression for
the vertex correction. The divergence in the vertex subdiagram is cancelled
by the counterterm depicted in (b) whereas the remaining overall divergence
in diagram (a) is cancelled by the counterter given in (c).

\newpage

\special{em:linewidth 0.4pt}
\unitlength 1mm
\linethickness{0.4pt}
\begin{picture}(125.33,30.00)
\put(7.17,17.50){\oval(3.00,25.00)[l]}
\put(9.67,25.00){\line(1,0){10.00}}
\put(28.33,25.00){\line(1,0){10.00}}
\put(24.00,25.00){\oval(8.00,6.00)[]}
\put(9.67,10.00){\line(1,0){10.00}}
\put(28.33,10.00){\line(1,0){10.00}}
\put(24.00,10.00){\oval(8.00,6.00)[]}
\put(41.50,17.50){\oval(3.00,25.00)[r]}
\put(45.00,29.00){\makebox(0,0)[lb]{-1}}
\put(55.17,17.50){\oval(6.33,7.00)[r]}
\put(55.00,14.00){\line(0,1){7.00}}
\put(55.00,21,00){\line(-2,1){6.00}}
\put(55.00,14.00){\line(-2,-1){6.00}}
\put(58.33,16.67){\line(1,0){6.67}}
\put(58.33,18.33){\line(1,0){6.67}}
\put(69.67,16.67){\line(1,0){4.67}}
\put(69.67,18.33){\line(1,0){4.67}}
\put(79.67,25.00){\line(1,0){26.00}}
\put(79.67,10.00){\line(1,0){26.00}}
\put(85.67,10.00){\line(0,1){15.00}}
\put(105.33,10.00){\line(0,1){15.00}}
\put(95.00,18.00){\makebox(0,0)[cc]{$V_{BS}$}}
\put(115.17,17.50){\oval(6.33,7.00)[r]}
\put(115.00,14.00){\line(0,1){7.00}}
\put(115.00,21.00){\line(-2,1){6.00}}
\put(115.00,14.00){\line(-2,-1){6.00}}
\put(118.33,16.67){\line(1,0){6.67}}
\put(118.33,18.33){\line(1,0){6.67}}
\end{picture}

\hspace*{7cm}{\bf a}
\vspace*{2.cm}

\special{em:linewidth 0.4pt}
\unitlength 1mm
\linethickness{0.4pt}
\begin{picture}(125.33,30.00)
\put(7.67,25.00){\line(1,0){19.67}}
\put(7.67,10.00){\line(1,0){19.67}}
\put(7.67,10.00){\line(0,1){15.00}}
\put(27.33,10.00){\line(0,1){15.00}}
\put(17.00,18.00){\makebox(0,0)[cc]{$V$}}
\put(31.67,16.67){\line(1,0){4.67}}
\put(31.67,18.33){\line(1,0){4.67}}
\put(39.67,25.00){\line(1,0){19.67}}
\put(39.67,10.00){\line(1,0){19.67}}
\put(39.67,10.00){\line(0,1){15.00}}
\put(59.33,10.00){\line(0,1){15.00}}
\put(51.00,18.00){\makebox(0,0)[cc]{$V_{BS}$}}
\put(59.33,25.00){\line(1,0){10.00}}
\put(78.00,25.00){\line(1,0){10.00}}
\put(73.67,25.00){\oval(8.00,6.00)[]}
\put(59.33,10.00){\line(1,0){10.00}}
\put(78.00,10.00){\line(1,0){10.00}}
\put(73.67,10.00){\oval(8.00,6.00)[]}
\put(93.17,17.50){\oval(3.00,25.00)[l]}
\put(95.67,25.00){\line(1,0){29.00}}
\put(95.67,10.00){\line(1,0){29.00}}
\put(127.50,17.50){\oval(3.00,25.00)[r]}
\put(131.00,29.00){\makebox(0,0)[lb]{-1}}
\end{picture}

\hspace*{7cm}{\bf b}
\vspace*{2.cm}

\special{em:linewidth 0.4pt}
\unitlength 1mm
\linethickness{0.4pt}
\begin{picture}(125.33,30.00)
\put(7.17,17.50){\oval(3.00,25.00)[l]}
\put(9.67,25.00){\line(1,0){29.00}}
\put(9.67,10.00){\line(1,0){29.00}}
\put(41.50,17.50){\oval(3.00,25.00)[r]}
\put(45.00,29.00){\makebox(0,0)[lb]{-1}}
\put(55.17,17.50){\oval(6.33,7.00)[r]}
\put(55.17,17.50){\oval(6.25,6.92)[r]}
\put(55.17,17.50){\oval(6.17,6.84)[r]}
\put(55.17,17.50){\oval(6.09,6.76)[r]}
\put(55.17,17.50){\oval(6.01,6.68)[r]}
\put(55.17,17.50){\oval(5.93,6.60)[r]}
\put(55.17,17.50){\oval(5.85,6.52)[r]}
\put(55.00,14.00){\line(0,1){7.00}}
\put(55.04,14.00){\line(0,1){7.00}}
\put(55.08,14.00){\line(0,1){7.00}}
\put(55.12,14.00){\line(0,1){7.00}}
\put(55.16,14.00){\line(0,1){7.00}}
\put(55.20,14.00){\line(0,1){7.00}}
\put(55.00,21,00){\line(-2,1){6.00}}
\put(55.00,14.00){\line(-2,-1){6.00}}
\put(58.33,16.67){\line(1,0){6.67}}
\put(58.33,18.33){\line(1,0){6.67}}
\put(69.67,16.67){\line(1,0){4.67}}
\put(69.67,18.33){\line(1,0){4.67}}
\put(79.67,25.00){\line(1,0){26.00}}
\put(79.67,10.00){\line(1,0){26.00}}
\put(85.67,10.00){\line(0,1){15.00}}
\put(105.33,10.00){\line(0,1){15.00}}
\put(95.00,18.00){\makebox(0,0)[cc]{$V$}}
\put(115.17,17.50){\oval(6.33,7.00)[r]}
\put(115.00,14.00){\line(0,1){7.00}}
\put(115.17,17.50){\oval(6.33,7.00)[r]}
\put(115.17,17.50){\oval(6.25,6.92)[r]}
\put(115.17,17.50){\oval(6.17,6.84)[r]}
\put(115.17,17.50){\oval(6.09,6.76)[r]}
\put(115.17,17.50){\oval(6.01,6.68)[r]}
\put(115.17,17.50){\oval(5.93,6.60)[r]}
\put(115.17,17.50){\oval(5.85,6.52)[r]}
\put(115.00,14.00){\line(0,1){7.00}}
\put(115.04,14.00){\line(0,1){7.00}}
\put(115.08,14.00){\line(0,1){7.00}}
\put(115.12,14.00){\line(0,1){7.00}}
\put(115.16,14.00){\line(0,1){7.00}}
\put(115.20,14.00){\line(0,1){7.00}}
\put(115.00,21.00){\line(-2,1){6.00}}
\put(115.00,14.00){\line(-2,-1){6.00}}
\put(118.33,16.67){\line(1,0){6.67}}
\put(118.33,18.33){\line(1,0){6.67}}
\end{picture}

\hspace*{7cm}{\bf c}
\vspace*{2.cm}

\centerline{\bf Fig. 1}
\newpage

\special{em:linewidth 0.4pt}
\unitlength 1.00mm
\linethickness{0.4pt}
\begin{picture}(146.00,19.00)
\put(10.00,18.00){\circle*{2.00}}
\put(10.00,17.00){\line(0,-1){1.33}}
\put(10.00,15.00){\line(0,-1){1.33}}
\put(10.00,13.00){\line(0,-1){1.33}}
\put(10.00,11.00){\line(0,-1){1.33}}
\put(10.00,9.00){\line(0,-1){1.33}}
\put(10.00,7.00){\circle*{2.00}}
\put(12.67,12.33){\line(1,0){2.33}}
\put(17.33,18.00){\circle*{2.00}}
\put(17.33,7.00){\circle*{2.00}}
\put(17.33,16.00){\circle*{0.00}}
\put(17.33,15.00){\circle*{0.00}}
\put(17.33,14.00){\circle*{0.00}}
\put(17.33,13.00){\circle*{0.00}}
\put(17.33,12.00){\circle*{0.00}}
\put(17.33,11.00){\circle*{0.00}}
\put(17.33,10.00){\circle*{0.00}}
\put(17.33,9.00){\circle*{0.00}}
\put(20.83,12.33){\oval(2.00,11.33)[r]}
\put(6.67,12.33){\oval(2.00,11.33)[l]}
\put(26.00,11.00){\line(0,1){2.67}}
\put(24.67,12.33){\line(1,0){2.67}}
\put(32.50,12.33){\oval(7.00,5.33)[]}
\put(35.33,14.33){\line(2,1){6.00}}
\put(35.33,10.00){\line(2,-1){6.00}}
\put(46.67,18.00){\circle*{2.00}}
\put(46.67,17.00){\line(0,-1){1.33}}
\put(46.67,15.00){\line(0,-1){1.33}}
\put(46.67,13.00){\line(0,-1){1.33}}
\put(46.67,11.00){\line(0,-1){1.33}}
\put(46.67,9.00){\line(0,-1){1.33}}
\put(46.67,7.00){\circle*{2.00}}
\put(49.33,12.33){\line(1,0){2.33}}
\put(54.00,18.00){\circle*{2.00}}
\put(54.00,7.00){\circle*{2.00}}
\put(54.00,16.00){\circle*{0.00}}
\put(54.00,15.00){\circle*{0.00}}
\put(54.00,14.00){\circle*{0.00}}
\put(54.00,13.00){\circle*{0.00}}
\put(54.00,12.00){\circle*{0.00}}
\put(54.00,11.00){\circle*{0.00}}
\put(54.00,10.00){\circle*{0.00}}
\put(54.00,9.00){\circle*{0.00}}
\put(57.50,12.33){\oval(2.00,11.33)[r]}
\put(43.34,12.33){\oval(2.00,11.33)[l]}
\put(62.33,11.00){\line(0,1){2.67}}
\put(61.00,12.33){\line(1,0){2.67}}
\put(70.33,18.00){\circle*{2.00}}
\put(70.33,17.00){\line(0,-1){1.33}}
\put(70.33,15.00){\line(0,-1){1.33}}
\put(70.33,13.00){\line(0,-1){1.33}}
\put(70.33,11.00){\line(0,-1){1.33}}
\put(70.33,9.00){\line(0,-1){1.33}}
\put(70.33,7.00){\circle*{2.00}}
\put(73.00,12.33){\line(1,0){2.33}}
\put(77.66,18.00){\circle*{2.00}}
\put(77.66,7.00){\circle*{2.00}}
\put(77.66,16.00){\circle*{0.00}}
\put(77.66,15.00){\circle*{0.00}}
\put(77.66,14.00){\circle*{0.00}}
\put(77.66,13.00){\circle*{0.00}}
\put(77.66,12.00){\circle*{0.00}}
\put(77.66,11.00){\circle*{0.00}}
\put(77.66,10.00){\circle*{0.00}}
\put(77.66,9.00){\circle*{0.00}}
\put(81.16,12.33){\oval(2.00,11.33)[r]}
\put(67.00,12.33){\oval(2.00,11.33)[l]}
\put(91.83,12.67){\oval(7.00,5.33)[]}
\put(89.00,14.67){\line(-2,1){6.00}}
\put(89.00,10.33){\line(-2,-1){6.00}}
\put(99.67,11.00){\line(0,1){2.67}}
\put(98.33,12.33){\line(1,0){2.67}}
\put(106.17,12.33){\oval(7.00,5.33)[]}
\put(109.00,14.33){\line(2,1){6.00}}
\put(109.00,10.00){\line(2,-1){6.00}}
\put(120.33,18.00){\circle*{2.00}}
\put(120.33,17.00){\line(0,-1){1.33}}
\put(120.33,15.00){\line(0,-1){1.33}}
\put(120.33,13.00){\line(0,-1){1.33}}
\put(120.33,11.00){\line(0,-1){1.33}}
\put(120.33,9.00){\line(0,-1){1.33}}
\put(120.33,7.00){\circle*{2.00}}
\put(123.00,12.33){\line(1,0){2.33}}
\put(127.66,18.00){\circle*{2.00}}
\put(127.66,7.00){\circle*{2.00}}
\put(127.66,16.00){\circle*{0.00}}
\put(127.66,15.00){\circle*{0.00}}
\put(127.66,14.00){\circle*{0.00}}
\put(127.66,13.00){\circle*{0.00}}
\put(127.66,12.00){\circle*{0.00}}
\put(127.66,11.00){\circle*{0.00}}
\put(127.66,10.00){\circle*{0.00}}
\put(127.66,9.00){\circle*{0.00}}
\put(131.16,12.33){\oval(2.00,11.33)[r]}
\put(117.00,12.33){\oval(2.00,11.33)[l]}
\put(142.50,12.67){\oval(7.00,5.33)[]}
\put(139.67,14.67){\line(-2,1){6.00}}
\put(139.67,10.33){\line(-2,-1){6.00}}
\put(32.50,12.33){\makebox(0,0)[cc]{$T_{12}$}}
\put(91.83,12.33){\makebox(0,0)[cc]{$T_{12}$}}
\put(106.17,12.33){\makebox(0,0)[cc]{$T_{12}$}}
\put(142.50,12.33){\makebox(0,0)[cc]{$T_{12}$}}
\put(13.82,0.00){\makebox(0,0)[cc]{{\bf a}}}
\put(46.67,0.00){\makebox(0,0)[cc]{{\bf b}}}
\put(77.67,0.00){\makebox(0,0)[cc]{{\bf c}}}
\put(124.15,0.00){\makebox(0,0)[cc]{{\bf d}}}
\end{picture}
\vspace*{2.cm}

\centerline{\bf Fig. 2}

\vspace*{3.cm}

\unitlength=1.00mm
\special{em:linewidth 0.4pt}
\linethickness{0.4pt}
\begin{picture}(139.00,23.00)
\put(14.00,14.40){\oval(8.00,8.00)[]}
\put(20.00,10.50){\oval(8.00,7.00)[lb]}
\put(20.00,18.20){\oval(8.00,7.00)[lt]}
\put(20.00,7.00){\line(1,0){9.00}}
\put(20.00,21.70){\line(1,0){9.00}}
\put(24.50,21.50){\oval(5.00,3.00)[b]}
\put(14.00,14.00){\makebox(0,0)[cc]{$T_{12}$}}
\put(32.00,14.00){\line(1,0){4.00}}
\put(34.00,12.00){\line(0,1){4.00}}
\put(43.00,14.40){\oval(8.00,8.00)[]}
\put(49.00,10.50){\oval(8.00,7.00)[lb]}
\put(49.00,18.20){\oval(8.00,7.00)[lt]}
\put(49.00,7.00){\line(1,0){9.00}}
\put(49.00,21.70){\line(1,0){9.00}}
\put(43.00,14.00){\makebox(0,0)[cc]{$T_{12}$}}
\put(53.50,7.10){\oval(5.00,3.00)[t]}
\put(62.00,14.00){\line(1,0){4.00}}
\put(64.00,12.00){\line(0,1){4.00}}
\put(73.00,14.40){\oval(8.00,8.00)[]}
\put(79.00,10.50){\oval(8.00,7.00)[lb]}
\put(79.00,18.20){\oval(8.00,7.00)[lt]}
\put(79.00,7.00){\line(1,0){10.00}}
\put(83.50,21.50){\oval(5.00,3.00)[b]}
\put(73.00,14.00){\makebox(0,0)[cc]{$T_{12}$}}
\put(89.00,18.20){\oval(8.00,7.00)[rt]}
\put(89.00,10.50){\oval(8.00,7.00)[rb]}
\put(79.00,21.70){\line(1,0){10.00}}
\put(95.00,14.40){\oval(8.00,8.00)[]}
\put(95.00,14.00){\makebox(0,0)[cc]{$T_{12}$}}
\put(102.00,14.00){\line(1,0){4.00}}
\put(104.00,12.00){\line(0,1){4.00}}
\put(113.00,14.40){\oval(8.00,8.00)[]}
\put(119.00,10.50){\oval(8.00,7.00)[lb]}
\put(119.00,18.20){\oval(8.00,7.00)[lt]}
\put(119.00,7.00){\line(1,0){10.00}}
\put(113.00,14.00){\makebox(0,0)[cc]{$T_{12}$}}
\put(129.00,18.20){\oval(8.00,7.00)[rt]}
\put(129.00,10.50){\oval(8.00,7.00)[rb]}
\put(119.00,21.70){\line(1,0){10.00}}
\put(135.00,14.40){\oval(8.00,8.00)[]}
\put(135.00,14.00){\makebox(0,0)[cc]{$T_{12}$}}
\put(123.50,7.10){\oval(5.00,3.00)[t]}
\put(23.82,0.00){\makebox(0,0)[cc]{{\bf a}}}
\put(51.67,0.00){\makebox(0,0)[cc]{{\bf b}}}
\put(84.50,0.00){\makebox(0,0)[cc]{{\bf c}}}
\put(124.15,0.00){\makebox(0,0)[cc]{{\bf d}}}
\end{picture}
\vspace*{2.cm}

\centerline{\bf Fig. 3}

\vspace*{3.cm}

\unitlength=1.00mm
\special{em:linewidth 0.4pt}
\linethickness{0.4pt}
\begin{picture}(42.00,23.00)
\put(10.00,22.00){\circle*{2.00}}
\put(10.00,7.00){\circle*{2.00}}
\put(10.00,14.50){\circle{6.00}}
\put(7.00,15.00){\vector(0,-1){1.00}}
\put(13.00,14.00){\vector(0,1){1.00}}
\put(10.00,20.00){\circle*{0.00}}
\put(10.00,19.00){\circle*{0.00}}
\put(10.00,10.00){\circle*{0.00}}
\put(10.00,9.00){\circle*{0.00}}
\put(16.00,14.50){\line(1,0){4.00}}
\put(18.00,12.50){\line(0,1){4.00}}
\put(39.00,22.00){\circle*{2.00}}
\put(39.00,7.00){\circle*{2.00}}
\put(39.00,14.50){\circle{6.00}}
\put(36.00,15.00){\vector(0,-1){1.00}}
\put(42.00,14.00){\vector(0,1){1.00}}
\put(39.00,20.00){\circle*{0.00}}
\put(39.00,19.00){\circle*{0.00}}
\put(39.00,10.00){\circle*{0.00}}
\put(39.00,9.00){\circle*{0.00}}
\put(26.50,14.50){\oval(9.00,8.00)[]}
\put(37.80,18.00){\oval(16.00,8.00)[lt]}
\put(37.80,11.00){\oval(16.00,8.00)[lb]}
\put(26.50,14.50){\makebox(0,0)[cc]{$T_{12}$}}
\put(45.00,14.50){\line(1,0){4.00}}
\put(47.00,12.50){\line(0,1){4.00}}
\put(55.00,22.00){\circle*{2.00}}
\put(55.00,7.00){\circle*{2.00}}
\put(55.00,14.50){\circle{6.00}}
\put(52.00,15.00){\vector(0,-1){1.00}}
\put(58.00,14.00){\vector(0,1){1.00}}
\put(55.00,20.00){\circle*{0.00}}
\put(55.00,19.00){\circle*{0.00}}
\put(55.00,10.00){\circle*{0.00}}
\put(55.00,9.00){\circle*{0.00}}
\put(67.50,14.50){\oval(9.00,8.00)[]}
\put(56.20,18.00){\oval(16.00,8.00)[rt]}
\put(56.20,11.00){\oval(16.00,8.00)[rb]}
\put(67.50,14.50){\makebox(0,0)[cc]{$T_{12}$}}
\put(75.00,14.50){\line(1,0){4.00}}
\put(77.00,12.50){\line(0,1){4.00}}
\put(86.50,14.50){\oval(9.00,8.00)[]}
\put(86.50,14.50){\makebox(0,0)[cc]{$T_{12}$}}
\put(99.00,22.00){\circle*{2.00}}
\put(99.00,7.00){\circle*{2.00}}
\put(99.00,14.50){\circle{6.00}}
\put(96.00,15.00){\vector(0,-1){1.00}}
\put(102.00,14.00){\vector(0,1){1.00}}
\put(99.00,20.00){\circle*{0.00}}
\put(99.00,19.00){\circle*{0.00}}
\put(99.00,10.00){\circle*{0.00}}
\put(99.00,9.00){\circle*{0.00}}
\put(97.80,18.00){\oval(16.00,8.00)[lt]}
\put(97.80,11.00){\oval(16.00,8.00)[lb]}
\put(111.50,14.50){\oval(9.00,8.00)[]}
\put(111.50,14.50){\makebox(0,0)[cc]{$T_{12}$}}
\put(100.20,18.00){\oval(16.00,8.00)[rt]}
\put(100.20,11.00){\oval(16.00,8.00)[rb]}
\put(10.00,0.00){\makebox(0,0)[cc]{{\bf a}}}
\put(32.80,0.00){\makebox(0,0)[cc]{{\bf b}}}
\put(60.33,0.00){\makebox(0,0)[cc]{{\bf c}}}
\put(99.00,0.00){\makebox(0,0)[cc]{{\bf d}}}
\end{picture}
\vspace*{2.cm}

\centerline{\bf Fig. 4}

\newpage

\unitlength=1.00mm
\special{em:linewidth 0.4pt}
\linethickness{0.4pt}
\begin{picture}(42.00,73.00)
\put(10.00,82.00){\circle*{2.00}}
\put(10.00,67.00){\circle*{2.00}}
\put(10.00,76.00){\circle*{2.00}}
\put(10.00,76.00){\circle*{0.00}}
\put(10.00,75.00){\circle*{0.00}}
\put(10.00,74.00){\circle*{0.00}}
\put(10.00,73.00){\circle*{0.00}}
\put(10.00,72.00){\circle*{0.00}}
\put(10.00,71.00){\circle*{0.00}}
\put(10.00,70.00){\circle*{0.00}}
\put(10.00,69.00){\circle*{0.00}}
\put(10.00,79.00){\circle{6.00}}
\put(16.00,74.50){\line(1,0){4.00}}
\put(18.00,72.50){\line(0,1){4.00}}
\put(39.00,82.00){\circle*{2.00}}
\put(39.00,67.00){\circle*{2.00}}
\put(39.00,76.00){\circle*{2.00}}
\put(39.00,76.00){\circle*{0.00}}
\put(39.00,75.00){\circle*{0.00}}
\put(39.00,74.00){\circle*{0.00}}
\put(39.00,73.00){\circle*{0.00}}
\put(39.00,72.00){\circle*{0.00}}
\put(39.00,71.00){\circle*{0.00}}
\put(39.00,70.00){\circle*{0.00}}
\put(39.00,69.00){\circle*{0.00}}
\put(39.00,79.00){\circle{6.00}}
\put(26.50,74.50){\oval(9.00,8.00)[]}
\put(37.80,78.00){\oval(16.00,8.00)[lt]}
\put(37.80,71.00){\oval(16.00,8.00)[lb]}
\put(26.50,74.50){\makebox(0,0)[cc]{$T_{12}$}}
\put(45.00,74.50){\line(1,0){4.00}}
\put(47.00,72.50){\line(0,1){4.00}}
\put(55.00,82.00){\circle*{2.00}}
\put(55.00,67.00){\circle*{2.00}}
\put(55.00,76.00){\circle*{2.00}}
\put(55.00,76.00){\circle*{0.00}}
\put(55.00,75.00){\circle*{0.00}}
\put(55.00,74.00){\circle*{0.00}}
\put(55.00,73.00){\circle*{0.00}}
\put(55.00,72.00){\circle*{0.00}}
\put(55.00,71.00){\circle*{0.00}}
\put(55.00,70.00){\circle*{0.00}}
\put(55.00,69.00){\circle*{0.00}}
\put(55.00,79.00){\circle{6.00}}
\put(67.50,74.50){\oval(9.00,8.00)[]}
\put(56.20,78.00){\oval(16.00,8.00)[rt]}
\put(56.20,71.00){\oval(16.00,8.00)[rb]}
\put(67.50,74.50){\makebox(0,0)[cc]{$T_{12}$}}
\put(75.00,74.50){\line(1,0){4.00}}
\put(77.00,72.50){\line(0,1){4.00}}
\put(86.50,74.50){\oval(9.00,8.00)[]}
\put(86.50,74.50){\makebox(0,0)[cc]{$T_{12}$}}
\put(99.00,82.00){\circle*{2.00}}
\put(99.00,67.00){\circle*{2.00}}
\put(99.00,76.00){\circle*{2.00}}
\put(99.00,76.00){\circle*{0.00}}
\put(99.00,75.00){\circle*{0.00}}
\put(99.00,74.00){\circle*{0.00}}
\put(99.00,73.00){\circle*{0.00}}
\put(99.00,72.00){\circle*{0.00}}
\put(99.00,71.00){\circle*{0.00}}
\put(99.00,70.00){\circle*{0.00}}
\put(99.00,69.00){\circle*{0.00}}
\put(97.80,78.00){\oval(16.00,8.00)[lt]}
\put(97.80,71.00){\oval(16.00,8.00)[lb]}
\put(99.00,79.00){\circle{6.00}}
\put(111.50,74.50){\oval(9.00,8.00)[]}
\put(111.50,74.50){\makebox(0,0)[cc]{$T_{12}$}}
\put(100.20,78.00){\oval(16.00,8.00)[rt]}
\put(100.20,71.00){\oval(16.00,8.00)[rb]}
\put(10.00,60.00){\makebox(0,0)[cc]{{\bf a}}}
\put(32.80,60.00){\makebox(0,0)[cc]{{\bf b}}}
\put(60.33,60.00){\makebox(0,0)[cc]{{\bf c}}}
\put(99.00,60.00){\makebox(0,0)[cc]{{\bf d}}}
\put(119.00,74.50){\line(1,0){4.00}}
\put(121.00,72.50){\line(0,1){4.00}}
\put(0.00,14.50){\line(1,0){4.00}}
\put(2.00,12.50){\line(0,1){4.00}}
\put(10.00,22.00){\circle*{2.00}}
\put(10.00,7.00){\circle*{2.00}}
\put(10.00,13.00){\circle*{2.00}}
\put(10.00,20.00){\circle*{0.00}}
\put(10.00,19.00){\circle*{0.00}}
\put(10.00,18.00){\circle*{0.00}}
\put(10.00,17.00){\circle*{0.00}}
\put(10.00,16.00){\circle*{0.00}}
\put(10.00,15.00){\circle*{0.00}}
\put(10.00,14.00){\circle*{0.00}}
\put(10.00,13.00){\circle*{0.00}}
\put(10.00,10.00){\circle{6.00}}
\put(16.00,14.50){\line(1,0){4.00}}
\put(18.00,12.50){\line(0,1){4.00}}
\put(39.00,22.00){\circle*{2.00}}
\put(39.00,7.00){\circle*{2.00}}
\put(39.00,13.00){\circle*{2.00}}
\put(39.00,20.00){\circle*{0.00}}
\put(39.00,19.00){\circle*{0.00}}
\put(39.00,18.00){\circle*{0.00}}
\put(39.00,17.00){\circle*{0.00}}
\put(39.00,16.00){\circle*{0.00}}
\put(39.00,15.00){\circle*{0.00}}
\put(39.00,14.00){\circle*{0.00}}
\put(39.00,13.00){\circle*{0.00}}
\put(39.00,10.00){\circle{6.00}}
\put(26.50,14.50){\oval(9.00,8.00)[]}
\put(37.80,18.00){\oval(16.00,8.00)[lt]}
\put(37.80,11.00){\oval(16.00,8.00)[lb]}
\put(26.50,14.50){\makebox(0,0)[cc]{$T_{12}$}}
\put(45.00,14.50){\line(1,0){4.00}}
\put(47.00,12.50){\line(0,1){4.00}}
\put(55.00,22.00){\circle*{2.00}}
\put(55.00,7.00){\circle*{2.00}}
\put(55.00,13.00){\circle*{2.00}}
\put(55.00,20.00){\circle*{0.00}}
\put(55.00,19.00){\circle*{0.00}}
\put(55.00,18.00){\circle*{0.00}}
\put(55.00,17.00){\circle*{0.00}}
\put(55.00,16.00){\circle*{0.00}}
\put(55.00,15.00){\circle*{0.00}}
\put(55.00,14.00){\circle*{0.00}}
\put(55.00,13.00){\circle*{0.00}}
\put(55.00,10.00){\circle{6.00}}
\put(67.50,14.50){\oval(9.00,8.00)[]}
\put(56.20,18.00){\oval(16.00,8.00)[rt]}
\put(56.20,11.00){\oval(16.00,8.00)[rb]}
\put(67.50,14.50){\makebox(0,0)[cc]{$T_{12}$}}
\put(75.00,14.50){\line(1,0){4.00}}
\put(77.00,12.50){\line(0,1){4.00}}
\put(86.50,14.50){\oval(9.00,8.00)[]}
\put(86.50,14.50){\makebox(0,0)[cc]{$T_{12}$}}
\put(99.00,22.00){\circle*{2.00}}
\put(99.00,7.00){\circle*{2.00}}
\put(99.00,13.00){\circle*{2.00}}
\put(99.00,20.00){\circle*{0.00}}
\put(99.00,19.00){\circle*{0.00}}
\put(99.00,18.00){\circle*{0.00}}
\put(99.00,17.00){\circle*{0.00}}
\put(99.00,16.00){\circle*{0.00}}
\put(99.00,15.00){\circle*{0.00}}
\put(99.00,14.00){\circle*{0.00}}
\put(99.00,13.00){\circle*{0.00}}
\put(97.80,18.00){\oval(16.00,8.00)[lt]}
\put(97.80,11.00){\oval(16.00,8.00)[lb]}
\put(99.00,10.00){\circle{6.00}}
\put(111.50,14.50){\oval(9.00,8.00)[]}
\put(111.50,14.50){\makebox(0,0)[cc]{$T_{12}$}}
\put(100.20,18.00){\oval(16.00,8.00)[rt]}
\put(100.20,11.00){\oval(16.00,8.00)[rb]}
\put(10.00,0.00){\makebox(0,0)[cc]{{\bf e}}}
\put(32.80,0.00){\makebox(0,0)[cc]{{\bf f}}}
\put(60.33,0.00){\makebox(0,0)[cc]{{\bf g}}}
\put(99.00,0.00){\makebox(0,0)[cc]{{\bf h}}}
\end{picture}
\vspace*{2.cm}

\centerline{\bf Fig. 5}

\vspace*{3.cm}

\unitlength=1.00mm
\special{em:linewidth 0.4pt}
\linethickness{0.4pt}
\begin{picture}(91.00,23.00)
\put(4.00,14.50){\oval(5.00,18.00)[l]}
\put(30.00,22.00){\circle*{2.00}}
\put(30.00,16.00){\circle*{2.00}}
\put(30.00,7.00){\circle*{2.00}}
\put(30.00,16.00){\circle*{0.00}}
\put(30.00,15.00){\circle*{0.00}}
\put(30.00,14.00){\circle*{0.00}}
\put(30.00,13.00){\circle*{0.00}}
\put(30.00,12.00){\circle*{0.00}}
\put(30.00,11.00){\circle*{0.00}}
\put(30.00,10.00){\circle*{0.00}}
\put(30.00,9.00){\circle*{0.00}}
\put(30.00,19.00){\circle{6}}
\put(15.00,14.50){\circle*{3.00}}
\put(29.00,14.50){\oval(28.00,15.00)[lt]}
\put(29.00,14.50){\oval(28.00,15.00)[lb]}
\put(15.00,14.50){\line(-1,1){8.00}}
\put(15.00,14.50){\line(-1,-1){8.00}}
\put(37.00,14.50){\line(1,0){4.00}}
\put(70.00,22.00){\circle{2.00}}
\put(70.00,22.00){\circle*{1.00}}
\put(70.00,7.00){\circle*{2.00}}
\put(70.00,20.00){\circle*{0.00}}
\put(70.00,19.00){\circle*{0.00}}
\put(70.00,18.00){\circle*{0.00}}
\put(70.00,17.00){\circle*{0.00}}
\put(70.00,16.00){\circle*{0.00}}
\put(70.00,15.00){\circle*{0.00}}
\put(70.00,14.00){\circle*{0.00}}
\put(70.00,13.00){\circle*{0.00}}
\put(70.00,12.00){\circle*{0.00}}
\put(70.00,11.00){\circle*{0.00}}
\put(70.00,19.00){\circle*{0.00}}
\put(70.00,10.00){\circle*{0.00}}
\put(70.00,9.00){\circle*{0.00}}
\put(55.00,14.50){\circle*{3.00}}
\put(69.00,14.50){\oval(28.00,15.00)[lt]}
\put(69.00,14.50){\oval(28.00,15.00)[lb]}
\put(55.00,14.50){\line(-1,1){8.00}}
\put(55.00,14.50){\line(-1,-1){8.00}}
\put(75.00,14.50){\oval(5.00,18.00)[r]}
\put(82.00,14.50){\line(1,0){4.00}}
\put(100.00,14.50){\circle{3.00}}
\put(100.00,14.50){\circle*{1.67}}
\put(101.20,15.70){\line(1,1){6.00}}
\put(101.20,13.30){\line(1,-1){6.00}}
\put(98.80,15.70){\line(-1,1){6.00}}
\put(98.80,13.30){\line(-1,-1){6.00}}
\put(15.00,0.00){\makebox(0,0)[cc]{{\bf a}}}
\put(55.00,0.00){\makebox(0,0)[cc]{{\bf b}}}
\put(100.00,0.00){\makebox(0,0)[cc]{{\bf c}}}
\end{picture}
\vspace*{2.cm}

\centerline{\bf Fig. 6}


\begin{thebibliography}{99}
\bibitem{ChPT} J. Gasser and H. Leutwyler, Ann. Phys. (N.Y.) {\bf 158}, 142 (1984).
\bibitem{Stern} M. Knecht, B. Moussalam, J. Stern and N. H. Fuchs,
                Nucl. Phys. {\bf B 457}, 513 (1995); {\bf B 471}, 445 (1996).
\bibitem{Bijnens} J. Bijnens et al., Phys. Lett. {\bf B 374}, 210 (1996).
\bibitem{Rosselet} L. Rosselet, {\it et al.}, Phys. Rev. D {\bf 15}, 574 (1977).
\bibitem{pNppN} E. A. Alekseeva {\it et al.}, Zh. Eksp. Teor. Fiz.
                {\bf 82}, 1007 (1982) [Sov. Phys. JETP {\bf 55}, 591 (1982)].
                O. O. Patarakin and V. N. Tikhonov, Kurchatov Institute of
                Atomic Energy preprint IAE-5629/2 (1993)
                D. Po\v{c}ani\'c, {\it et al.}, Phys. Rev. Lett.
                {\bf 72}, 1156 (1994).
                H.~Burkhardt and J. Lowe, Phys. Rev. Lett. {\bf 67}, 2622 (1991).
\bibitem{Pocanic} D. Po\v{c}ani\'{c},
                  Contribution to Proceedings of Workshop on Chiral Dynamics,
                  Theory and Experiment, Massachusetts Institute of
                  Technology, July 25 - 29, 1994, Eds. A.~Bernstein,
                  B. Holstein;
                  Summary of the working group on $\pi\pi$ and $\pi N$
                  interactions given at Workshop on Chiral Dynamics:
                  Theory and Experiment (ChPT 97), Mainz, Germany,
                  1-5 September 1997 (in print), hep-ph/9711361.
\bibitem{Nemenov} L. L. Nemenov, Sov. J. Nucl. Phys. {\bf 41}, 629 (1985).
\bibitem{Afanasyev} L.G. Afanasyev et al., Phys. Lett. {\bf B308}, 200 (1993);
                    {\it ibid} {\bf B338}, 478 (1994).
\bibitem{Deser} S. Deser {\it et al.}, Phys. Rev. {\bf 96}, 774 (1954).
\bibitem{Uretsky} J. L. Uretsky and T. R. Palfrey, Phys. Rev. {\bf 121}, 1798 (1961).
                  S. M. Bilenky {\it et al.}, Yad.~Fiz. {\bf 10}, 812 (1969).
\bibitem{Sigg} D. Sigg {\it et al.}, Phys. Rev. Lett. {\bf 75}, 3245 (1995).
\bibitem{Chattelard} D. Chattelard {\it et al.}, Phys. Rev. Lett. {\bf 74}, 4157 (1995).
\bibitem{Trueman} T. L. Trueman, Nucl. Phys. {\bf 26}, 57 (1961).
\bibitem{Rasche1} U. Moor, G. Rasche and W.S. Woolcock, Nucl. Phys. {\bf A587}, 747 (1995).
\bibitem{Rasche2} A. Gashi, G.~Rasche, G. C. Oades and W. S. Woolcock, Nucl. Phys. {\bf A628}, 101 (1998)
\bibitem{Gibbs} P. B. Siegel and W. R. Gibbs, Phys. Rev. {\bf C33}, 1407 (1986);
                W. B. Kaufmann and W.~R.~Gibbs, Phys. Rev. {\bf C35}, 838 (1987).
\bibitem{Geramb} M. Sander, C. Kuhrts and H. V. von Geramb, Phys. Rev. {\bf C53}, R2610 (1996).
\bibitem{Efimov} G. V. Efimov, M. A. Ivanov and V. E. Lyubovitskij,
                 Sov. J. Nucl. Phys. {\bf 44}, 296 (1986).
\bibitem{Pervushin} A. A. Bel'kov, V. N. Pervushin and F. G. Tkebuchava,
                 Sov. J. Nucl. Phys. {\bf 44}, 300 (1986).
\bibitem{Silagadze} Z. Silagadze, JETP Lett., {\bf 60}, 689 (1994).
\bibitem{Kuraev} E. A. Kuraev, Phys. At. Nucl. {\bf 61}, 239 (1998).
\bibitem{Sazdjian} H. Jallouli and H. Sazdjian, Preprint IPNO/TH 97-01 (1997).
\bibitem{Labelle} P. Labelle and K. Buckley, Preprint hep-ph/9804201.
\bibitem{PLB} V. E. Lyubovitskij and A. G. Rusetsky, Phys. Lett. {\bf B389}, 181 (1996).
\bibitem{JETP} V. E. Lyubovitskij, E. Z. Lipartia and A. G. Rusetsky,
               JETP Lett., {\bf 66}, 747 (1997).
\bibitem{Barbieri} R. Barbieri and E. Remiddi, Nucl. Phys. {\bf B141}, 413 (1978).
\bibitem{Schwinger} J. Schwinger, J. Math. Phys. {\bf 5}, 1606 (1964).
\bibitem{Goldberger} M. L. Goldberger and K. M. Watson, Collision Theory,
                     J. Wiley and Sons, New-York -- London -- Sydnay (1964).
\bibitem{Roig} F. S. Roig and A. R. Swift, Nucl. Phys. {\bf B104}, 548 (1976).
\bibitem{Schweber} S. Schweber, An Introduction to Relativistic Quantum
                   Field Theory, Row, Peterson and Co, Evanston Ill.,
                   N.Y., 1961, Chap. 15.
\bibitem{Knecht} M. Knecht and R. Urech, Preprint hep-ph/9709348.
\end{thebibliography}
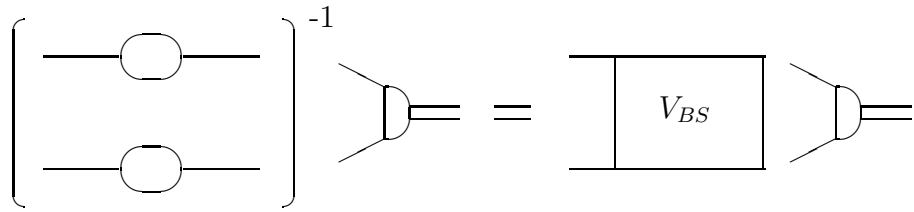
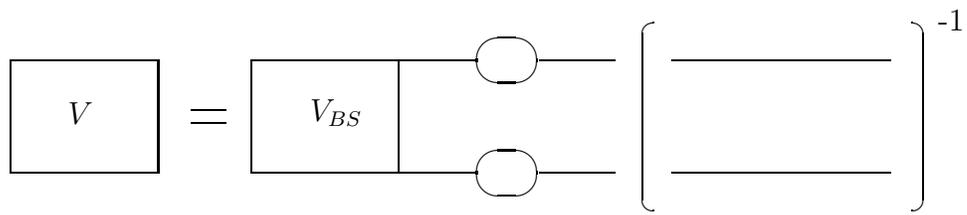
\end{document}